\documentclass[prd,twocolumn,longbibliography]{revtex4-1}

\usepackage{epsfig,amsmath,amssymb}
\usepackage[outdir=./]{epstopdf}

\newcommand{\bra}[1]{\langle #1|}
\newcommand{\ket}[1]{|#1\rangle}
\newcommand{\Exp}[1]{\langle#1\rangle}

\begin{document}
	
\title{Dual to the anomalous weak value effect of photon-polarisation separation}

\author{James Q. Quach} 
\email{quach.james@gmail.com}
\affiliation{Institute for Photonics and Advanced Sensing and School of Chemistry and Physics, The University of Adelaide, South Australia 5005, Australia}

\begin{abstract}
	The quantum Cheshire cat (QCC) thought experiment proposes that a quantum object's property (\textit{e.g} polarisation, spin, etc.) can be separated from its physical body or \textit{disembodied}. This conclusion arose from an argument that interprets a zero weak value (WV) of polarisation as no polarisation. We show that this argument is incomplete in the sense that a zero WV reading could equally be interpreted as linear polarisation. Nevertheless, through a generalisation of the QCC, we complete their argument by excluding the possibility of linear polarisation as a consistent interpretation.  We go further, and introduce the dual of the generalised QCC. The dual QCC exhibits an intriguing effect, where a horizontally-polarised interferometer with just one arm, can give rise to interference which is vertically-polarised. The interference appears to arise as the result of the phase difference between the physical arm and a \textit{phantom} arm. This peculiar effect arises from the interplay between the pre-selected and post-selected states, which characterises WVs. The QCC has not yet been unambiguously experimentally demonstrated. The QCC dual offers an alternative pathway to experimental realisation.
\end{abstract}
	
\maketitle

\section{Introduction}
\label{sec:Introduction} 
The conventional view of measurement in quantum mechanics is that it is a destructive process that irrevocably projects the system into an eigenstate of the observed variable. \textit{Weak measurements }provides a formal non-destructive measurement scheme by weakly coupling the system to an ancilla, and performing a measurement (projection) on the ancilla in some appropriate basis~\cite{aharonov88}. Operationally, the ancilla is a measurement device with a pointer; the interaction of the system and the ancilla shifts the pointer state proportional to the magnitude of the observed variable. As the ancilla interacts only very weakly with the system, the state can evolve without appreciable disturbance.

\textit{Weak values} seek to represent the observables of intermediate states, as the system evolves from a pre-selection to a post-selection state. It is a unique consequence of quantum mechanics that one may choose both pre-selection \textit{and} post-selection states, which distinguishes it from classical mechanics, where the choice of the initial state defines the final state, or vice versa. This idea is more generally explored in the two-state vector formalism~\cite{aharonov02}. By judicious post-selection, WVs have been used to amplify small signals~\cite{brunner10,dixon09,egan12,feizpour11,hogan11,hosten08,jayaswal14,jordan14,knee14,pfeifer11,starling09,starling09,starling10,starling10a,strubi13,turner11,viza13,zhou12,zhou13}, provide direct determination of quantum states and geometric phases via the complex nature of WVs~\cite{kobayashi10,kobayashi11,lundeen09,lundeen11,lundeen14,malik2014,salvail13,sjoqvist06}, and give conditioned averages associated with observables~\cite{brunner04,mir07,kocsis11}. In an intriguing proposal Aharanov \textit{et al.}~\cite{aharonov13} showed that WVs can give rise to a situation where the position of a photon exists in one arm of an interferometer, whilst its polarisation exists in the other arm. The effect was given the name quantum Cheshire cat, which alludes to Lewis Carroll's Cheshire cat, whose grin (polarisation) could exist without its body (photon). 

The search for dualities has been a fruitful path to insights and novel phenomena in physics; \textit{e.g.} the wave-particle duality, electromagnetic duality, the Aharanov-Casher effect~\cite{casher84} and its dual~\cite{he93,wilkens94}, and many more. Here we first generalise the QCC with elliptical polarisations, and then we introduce the dual of the QCC, and show its novel behaviour. There has been a discussion on the physical interpretation of WVs, since it's inception (see \cite{aharonov08,aharonov10,parrott09,svensson13,svensson13a} and references therein). We take the approach that a WV represents a physical property of the quantum system being measured, in the same spirit as the original QCC.

%In Sec.~\ref{sec:WVs} and \ref{sec:Quantum Cheshire Cat} we review WVs and the QCC. In Sec.~\ref{sec:Slit} we present the dual of the QCC. In Sec.~\ref{sec:double-slit experiment} we discuss temporal interference. 

\section{Weak values}
\label{sec:WVs} 
If we precisely know the position of a quantum particle, we have no information about its speed. However we may place weak detectors all around the particle to deduce its average speed, by measuring the time it took to reach the detectors. We may also ask, what is the speed of the particle to reach a subset of locations, as detected by a subensemble of the detectors? The answer to this question is a WV.

Prior to measurement the pre-selected state $\ket{\psi_i}$ and pointer state $\ket{m_i}$ are uncoupled. In the weak measurement scheme, the interaction Hamiltonian between the system and pointer is
\begin{equation}
	\hat{H}_{\text{int}} = g(t)\hat{O}\hat{P}~,
\end{equation}
which couples the system's observable $\hat{O}$ to the pointer momentum $\hat{P}$. The interaction with the pointer occurs for a short time, outside of which coupling constant $g$ is zero, so that the evolutionary operator is $\hat{U} = \exp(-\frac{i}{\hbar}\int\hat{H}_\text{int}dt) = \exp(-\frac{i}{\hbar}g\hat{O}\hat{P})$. After the interaction with the pointer, the system undergoes a projective measurement where only a subset of the measured states are chosen. Labelling this post-selected state $\ket{\psi_f}$, the final pointer state is ($\hbar=1$)
\begin{equation}
\begin{split}
	\ket{m_f} &= \bra{\psi_f}\exp(-ig\hat{O}\hat{P})\ket{\psi_i}\ket{m_i}\\
		&\approx \langle \psi_f |\psi_i\rangle\Bigl(1 - ig\frac{\bra{\psi_f}\hat{O}\ket{\psi_i}}{\langle \psi_f |\psi_i\rangle }\hat{P}\Bigr)\ket{m_i}\\
		&\approx \langle \psi_f |\psi_i\rangle\exp(-ig\Exp{\hat{O}}_w\hat{P})\ket{m_i}
\end{split} 
\end{equation}
where
\begin{equation}
	\Exp{\hat{O}}_w \equiv \frac{\bra{\psi_f}\hat{O}\ket{\psi_i}}{\langle \psi_f |\psi_i\rangle }
\label{eq:weak}
\end{equation}
is known as the WV of $\hat{O}$.

The pointer momentum $\hat{P}$ is conjugate to the pointer position $\hat{X}$. Let us now write the initial pointer state in the position basis, $\ket{m_i} = \int dx\ket{x}\varphi(x)$, where $\varphi(x)\equiv \Exp{x|m_i}$ and is assumed to be real. The final pointer state in the position basis then is
\begin{equation}
\begin{split}
	\ket{m_f} &\approx \langle \psi_f |\psi_i\rangle\exp(-ig\Exp{\hat{O}}_w\hat{P})\int dx\ket{x}\varphi(x)\\
		&= \langle \psi_f |\psi_i\rangle  \int dx\ket{x}\varphi(x - g\Exp{\hat{O}}_w)~,
\label{eq:mf}
\end{split} 
\end{equation}
where we have used the fact that $\hat{P}$ acts as a translation operator that shifts the pointer state in the conjugate $x$-basis by $g\Exp{\hat{O}}_w$. If the pointer states were the positions of a needle on a measuring device, the interaction of the measurement device with the system will shift the position of the needle by a distance proportional to $\Exp{\hat{O}}_w$, thereby giving us a measurement of observable $\hat{O}$. 

\section{Generalised Quantum Cheshire Cat}
\label{sec:Quantum Cheshire Cat}

\begin{figure}
	\centering
	\includegraphics[width=.5\columnwidth]{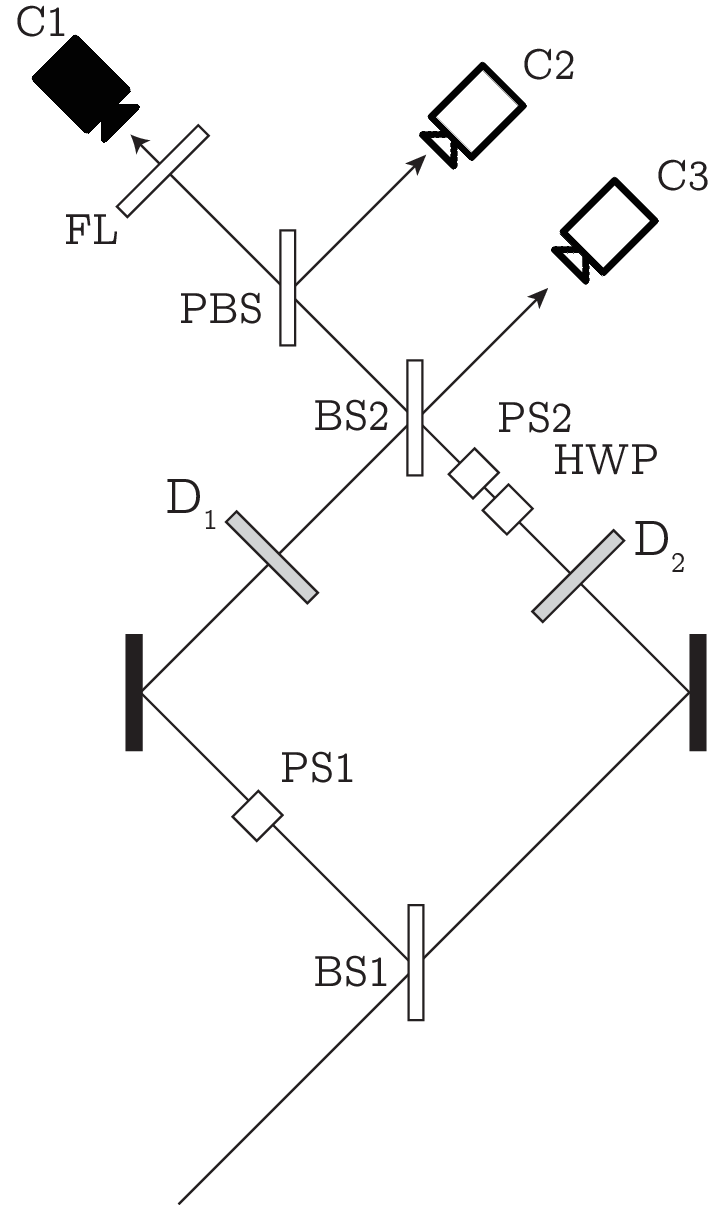}
	\caption{A schematic of the the generalised QCC, where a phase-shifter (PS1) introduces a controllable phase difference $e^{i\theta}$ between the two arms.  A half-wave plate (HWP), phase-shifter (PS2), beam-splitter (BS2), and polarising beam-splitter (PBS), are used for post-selection. To project the pointer onto the momentum basis, a Fourier lens (FL) is used to Fourier transform the light beam so that each pixel on the camera corresponds to a transverse momentum. To project the pointer onto the position basis, remove the FL, and each pixel corresponds to a transverse displacement. Detectors $\hat{\sigma}_A$ and $\hat{A}$ are placed at location $D_1$, and $\hat{\sigma}_B$ and $\hat{B}$ at location $D_2$.}
	\label{fig:phantom}
\end{figure}

The QCC is an interferometer experiment with pre-selection, post-selection, and weak detectors. We generalise the pre-selection state of the QCC with a phase differential between the arms of the interferometer,  
\begin{equation}
	\ket{\Phi_i} = (e^{i\theta}\ket{A} + \ket{B})\ket{H}/\sqrt{2}~,
\label{eq:QCC_in}
\end{equation}
where $\ket{A} (\ket{B})$ represents a state located in arm A (B) of the interferometer, and $\ket{H} (\ket{V})$ is horizontal (vertical) polarisation. The phase differential can be implemented with a phase-shifter (PS1) in one of the arms as shown in Fig~\ref{fig:phantom}. The original QCC pre-selection state is a special case of Eq.~(\ref{eq:QCC_in}), where $\theta=\pi/2$.

The QCC post-selection state is 
\begin{equation}
	\ket{\Phi_f} = (\ket{A}\ket{H} + \ket{B}\ket{V})/\sqrt{2}~.
\label{eq:QCC_out}
\end{equation}

The projectors in the QCC experiment measure the presence of photons in arms $A$ and $B$,
\begin{align}
	\hat{A}=\ket{A}\bra{A} (\ket{L}\bra{L} + \ket{R}\bra{R})~,\label{eq:A}\\
	\hat{B}=\ket{B}\bra{B} (\ket{L}\bra{L} + \ket{R}\bra{R})~,\label{eq:B}
\end{align}
and the polarisation in arms $A$ and $B$,
\begin{align}
\hat{\sigma}_A = \ket{A}\bra{A}(\ket{L}\bra{L} - \ket{R}\bra{R})~,\label{eq:sigmaA}\\
\hat{\sigma}_B = \ket{B}\bra{B}(\ket{L}\bra{L} - \ket{R}\bra{R})~,\label{eq:sigmaB}
\end{align}
where 
\begin{gather}
	\ket{L}\equiv (\ket{H}+e^{i\phi}\ket{V})/\sqrt{2}~,\label{eq:L}\\
	\ket{R}\equiv (\ket{H}-e^{i\phi}\ket{V})/\sqrt{2}~\label{eq:R}
\end{gather}
are left-elliptical and right-elliptical polarisation states. In comparison to the original QCC experiment, we have generalised the basis states to elliptical polarisation states with the phase parameter $\phi$. The original QCC circular polarisation basis is a special case of Eq.~(\ref{eq:L}) and (\ref{eq:R}), where $\phi=\pi/2$.

$\hat{A}$ ($\hat{B}$) detects whether there is a photon in arm A (B). $\hat{\sigma}_A$ detects the polarisation of the photon in arm A. The eigenvalues 1 and -1 correspond to eigenstates $\ket{L}\ket{A}\equiv\ket{L,A}$ and $\ket{R,A}$, which are states of left-elliptical and right-elliptical polarised photons in arm A respectively; whereas the eigenvalue 0 corresponds to the degenerate subspace spanned by eigenstates $\ket{L,B}$ and $\ket{R,B}$, which are states of photons in arm B. $\hat{\sigma}_B$ is similarly defined.  

Using Eq.~(\ref{eq:weak}), the WVs measured by these operators are
\begin{align}
	\Exp{\hat{A}}_w =\Exp{\hat{A}\hat{L}}_w + \Exp{\hat{A}\hat{R}}_w= 1~,\label{eq:hatA}\\
	\Exp{\hat{B}}_w =\Exp{\hat{B}\hat{L}}_w + \Exp{\hat{B}\hat{R}}_w= 0~\label{eq:hatB}
\end{align} 
and
\begin{align}
\Exp{\hat{\sigma}_A}_w &=\Exp{\hat{A}\hat{L}}_w - \Exp{\hat{A}\hat{R}}_w= 0~,\label{eq:hatSigmaA}\\
\Exp{\hat{\sigma}_B}_w &= \Exp{\hat{B}\hat{L}}_w - \Exp{\hat{B}\hat{R}}_w=e^{i(\phi-\theta)}~,\label{eq:hatSigmaB}
\end{align} 
where
\begin{align}
	\hat{L}&=\ket{L}\bra{L} (\ket{A}\bra{A} + \ket{B}\bra{B})~,\\
	\hat{R}&=\ket{R}\bra{R} (\ket{A}\bra{A} + \ket{B}\bra{B})~.
\end{align} 
Note that $\Exp{\hat{A}\hat{L}}_w=\Exp{\hat{A}\hat{R}}_w=1/2$ and $\Exp{\hat{B}\hat{L}}_w=-\Exp{\hat{B}\hat{R}}_w=e^{i(\phi-\theta)}/2$. For $\theta=\phi=\pi/2$ we retrieve the original QCC result, which the QCC authors interpret as the photon existing in arm A, whilst its left-circular polarisation is detected in arm B. The generalised QCC generalises this to the detection of elliptical polarisation in arm B. It shows that the polarisation is determined by the phase difference between the interferometer arms. Specifically, let us rotate the polarisation basis of $\hat{\sigma}_B$ so that $\phi=\theta$. In this basis $\Exp{\hat{\sigma}_B}_w=1$. Generalising the QCC, we interpret this to mean that arm B has an elliptical polarisation that is dependent on the phase difference between the interferometer arms.

\subsection{Interpretation}
\label{sec:Interpretation}

The WV, defined relative to a given observable and a pair of pre- and post-selected state [Eq.~(\ref{eq:weak})], is formally unambiguous as it lies within the standard quantum mechanical framework. It is the interpretation of the WV that is controversial. The reason for this is that traditionally quantum mechanics assigns properties to systems only upon projective measurements. Projective measurements however, necessarily alters the quantum system's state; weak measurements seeks to understand system properties in a non-destructive way.

Currently, interpretations of the WV can be categorised into three camps:
\begin{itemize}
	\item WVs are numbers stemming from perturbation theory that have no relation to system properties~\cite{sokolovski15},
	\item WVs are pre- and post-selected ensemble averages (conditional expectation values), but do not represent \textit{genuine} system properties~\cite{svensson13},
	\item WVs partially represent local properties of the system for a given pair of pre- and post-selected in a retrodictive manner~\cite{matzkin19}.
\end{itemize}

The QCC argument would fall in the third interpretative category. As such, we will adopt the interpretation that WVs do represent, at least partially, local properties of the system.

In the QCC it is implicitly assumed that a $\Exp{\hat{\sigma}_A}_w = 0$, corresponds to no polarisation detected in arm A, which was not completely justified in the original paper. Let us review what occurs in a weak measurement. In a weak measurement, a measurement device weakly couples to the degree of freedom that one wishes to measure, \textit{e.g.} a particle's polarisation. After the interaction with the weak measurement device, one destructively measures the state of the system. If the final state of the system corresponds to some predefined post-selected state, then one records the reading on the weak measurement device, otherwise one ignores the reading.

Now if one wishes to interpret a reading on the measurement device as a measure of a property of the system, the interpretation of this value should be independent of the pre- and post-selected states. Most notably, if one chooses the post-selected state to be the same as pre-selected state, then the WV is simply the expectation value, \textit{i.e.} expectation values are a subset of weak values. The interpretation of the reading on the measurement device should be consistent with the interpretation of the reading for the expectation value.
Specifically, suppose we have a circular-polarisation and a linear-polarisation detector, $\hat{\sigma}_A^{C}$ and $\hat{\sigma}_A^{L}$. If $\Exp{\hat{\sigma}_A^C}_w=\bra{\psi_i}\hat{\sigma}_A^C\ket{\psi_i}_w = 1$ and -1, this should be interpreted to mean that the particle is left- and right-circular polarised, respectively. Similarly, $\Exp{\sigma_A^L}_w=\bra{\psi_i}\hat{\sigma}_A^L\ket{\psi_i}_w = 1$ and -1, should be interpreted to mean that the particle is horizontally and vertically polarised, respectively. Now, when $\Exp{\hat{\sigma}_A^C}_w = 0$ this should be interpreted as no \textit{circular} polarisation, not necessarily no polarisation, as claimed in the original QCC paper. The reason for this is illustrated when $\ket{\psi_i}=(\ket{H} + i\ket{V})\ket{A}/\sqrt{2}$: here $\Exp{\hat{\sigma}_A^C}_w = 0$ but $\Exp{\hat{\sigma}_A^L}_w = 1$. In this case, the only consistent interpretation is that the particle is horizontally polarised. Building on this insight, we use the generalised QCC to complete the original QCC argument in a consistent manner. 

Consider the case when $\phi-\theta=\pi/2$: $\Exp{\hat{\sigma}_A}_w = 0$ and $\Exp{\hat{\sigma}_B}_w = i$. For both operators, the WV is zero, as read by the expected value of the pointer \textit{i.e.} $\text{Re}\Exp{\hat{\sigma}_A}_w=\text{Re}\Exp{\hat{\sigma}_B}_w=0$ (it is only the real component of WVs that shifts the pointer state~\cite{aharonov88}). Let us now rotate the basis of the polarisation detector so that $\phi=\theta$: $\Exp{\hat{\sigma}_A}_w = 0$ and $\Exp{\hat{\sigma}_B}_w = 1$. Now in arm B, the measured WV is no longer 0; in comparison, the WV of polarisation in arm A is still 0.  In fact, it does not matter how we rotate the basis of the polarisation operator, $\text{Re}\Exp{\hat{\sigma}_A}_w$ will always be 0; whereas $\text{Re}\Exp{\hat{\sigma}_B}_w=\cos(\phi-\theta)$ is in general non-zero.

Let us compare this with reading for the expectation value for a known polarised particle in arm B: $\ket{\psi_i}=(\ket{H}+e^{i\theta'}\ket{V})\ket{B}/\sqrt{2}$. The average reading on the measurement device is $\Exp{\hat{\sigma}_B}=\cos(\phi-\theta')$, which exactly corresponds to the weak value reading $\text{Re}\Exp{\hat{\sigma}_B}_w$. In other words, a zero WV reading in the polarisation pointer for arm B is polarisation basis dependent, whereas for arm A the zero WV is basis independent. As $\Exp{\hat{\sigma}_A}_w$ always vanishes no matter on which polarisation basis we measure, the generalised QCC supports the idea that $\Exp{\hat{\sigma}_A}_w = 0$ should be interpreted as no polarisation. This interpretation is also consistent with the expectation value, where there is no polarisation only if $\bra{\psi_i}\hat{\sigma}_A\ket{\psi_i}_w=0~\forall \phi$, otherwise there is polarisation.

There is another degree of freedom that we have not discussed, which is the choice of the spatial co-ordinate axes. Including this degree of freedom, the detector polarisation basis is 
\begin{gather}
\ket{L}\equiv (\cos\chi\ket{H}+e^{i\phi}\sin\chi\ket{V})/\sqrt{2}~,\label{eq:Lchi}\\
\ket{R}\equiv (\sin\chi\ket{H}-e^{i\phi}\cos\chi\ket{V})/\sqrt{2}~,\label{eq:Rchi}
\end{gather}
where $\chi$ is the co-ordinate axes angle of rotation. This leads to
\begin{align}
	\Exp{\hat{\sigma}_A}_w &= 0~,\label{eq:hatSigmaAchi}\\
	\Exp{\hat{\sigma}_B}_w &= 2\cos\chi\sin\chi e^{i(\phi-\theta)}~.\label{eq:hatSigmaBchi}
\end{align}

For $\chi=0$ or $\pi/2$, $\Exp{\hat{\sigma}_B}_w=0$. To determine whether this zero value is an artefact of the co-ordinate choice, one should rotate the co-ordinate axes of the polarisation detectors. Note that to vary $\chi$ one can simply rotate the birefringent crystal that implements the polarisation detector, whereas to vary $\phi$ one would need a different crystal with different refractive properties (Sec.~\ref{QCC Implementation}).  If $\Exp{\hat{\sigma}_B}_w$ is no longer zero after rotating the polarisation detector, then one should attribute the zero reading to the choice of co-ordinate. This is consistent with the expectation value, which in arbitrary co-ordinate rotation is $\bra{\psi_i}\hat{\sigma}_B\ket{\psi_i}_w = 2\cos\chi\sin\chi \cos(\phi-\theta')$. For the rest of this paper, we will use co-ordinates where $\chi=\pi/4$.

\subsection{Probability of detection}
\label{sec:Probability of detection}

If there is no polarisation in arm A, then a polarisation detector should not interact with the system and therefore does not disturb it in anyway. In contrast, the detection of polarisation would necessarily disturb the system, affecting the probability of detection. Without interaction the probability of detection is the overlap between the pre-selected and post-selected states; projected onto pointer basis $q$ this is
\begin{equation}
\mathcal{P}=|\Exp{\psi_f|\psi_i}|^2|\Exp{q|m_i}|^2~.
\end{equation}

In general, the probability of detection after interaction with the detector is 
\begin{equation}
\mathcal{P}_\epsilon=|\Exp{q|\bra{\psi_f}\hat{U}\ket{m_i}|\psi_i}|^2~,
\end{equation}
where $\hat{U} = \exp(-\frac{i}{\hbar}g\hat{O}\hat{P})$.

To first order in $g$ this gives~\cite{dressel14a}
\begin{equation}
\frac{\mathcal{P}_\epsilon}{\mathcal{P}}-1=\frac{2g}{\hbar}\Big(\text{Re}\Exp{\hat{O}}_w\text{Im}\Exp{\hat{P}}_w+\text{Im}\Exp{\hat{O}}_w\text{Re}\Exp{\hat{P}}_w\Big)~,
\label{eq:PRatio}
\end{equation}
where
\begin{equation}
\Exp{\hat{P}}_w = \frac{\bra{q}\hat{P}\ket{m_i}}{\langle q |m_i\rangle }~.
\label{eq:Pw}
\end{equation}
$\Exp{\hat{P}}_w$ is the momentum WV of the pointer, which is dependent on the choice of basis. In Sec.~\ref{QCC Implementation} we give specific examples.

Consider again the case $\Exp{\hat{\sigma}_A}_w = 0$ and $\Exp{\hat{\sigma}_B}_w = i$. From Eq.~(\ref{eq:PRatio}), we see no disturbance in the probability of detection in $\Exp{\hat{\sigma}_A}_w=0$, $\mathcal{P}_\epsilon=\mathcal{P}$; whereas for $\Exp{\hat{\sigma}_B}_w=i$, there is a change in the probability of detection given by, 
\begin{equation}
\frac{\mathcal{P}_\epsilon}{\mathcal{P}}-1=\frac{2g}{\hbar}\text{Re}\Exp{\hat{P}}_w~.
\label{eq:PRatioSimple}
\end{equation}
In other words, the $\hat{\sigma}_B$ operator disturbs the probability of detection by an amount proportional to the momentum WV of the pointer.

In fact this is true no matter what elliptical basis we choose to measure the polarisation in. As $\Exp{\hat{\sigma}_A}_w=0$ always vanishes, the probability of detection is indistinguishable from no measurement; whereas $\Exp{\hat{\sigma}_B}_w = e^{i(\phi-\theta)}$ will always disturb the probability of detection (for $\Exp{\hat{P}}_w\neq 0$). As $\hat{\sigma}_A$ does not disturb the probability of detection and does not shift the polarisation pointer, we identify $\Exp{\hat{\sigma}_A}_w = 0$ as corresponding to no polarisation. 

The interpretation that $\Exp{\hat{\sigma}_A}_w = 0$ corresponds to no polarisation rests on the epistemic assumptions that WVs correspond to physical properties of a system, and that these values are consistent with standard expectation values. Therefore, our interpretation is only as valid as the strength of these assumptions, which belong to the wider interpretive issue of WVs in general. However, adopting these assumptions has allowed us to extend the analysis of the QCC in a consistent manner.

\subsection{Implementation}
\label{QCC Implementation}

As a \textit{gedanken} experiment, the authors of the QCC considered an interferometer setup with a series of optical elements and detectors for post-selection,
as laid-out in Fig.~\ref{fig:phantom}. Post-selection is achieved with a half-wave plate (HWP), phase-shifter (PS2), beam-splitter (BS2), and polarising beam-splitter (PBS). The HWP flips  polarisation $\ket{H}\leftrightarrow\ket{V}$. PS2 shifts the phase by $i$. The PBS transmits horizontal polarisation and reflects vertical polarisation. Under this construction, states orthogonal to $\ket{\Phi_f}$ will not trigger detector C1 (they will trigger C2 or C3), and $\ket{\Phi_f}$ will trigger C1 with certainty. Post-selection means that we will only consider measurements that coincide with the triggering of C1. 

The $\hat{A}$ detector could be implemented with a sheet of glass placed in arm A [position D1 in Fig.~\ref{fig:phantom}(a)], slightly tilted up to produce a small vertical displacement of the beam. The C1 detector could be a CCD camera to record the beam deflection. Detection of the deflection would indicate the photon went via arm A. Similarly, the $\hat{B}$ detector could be implemented with a sheet of glass placed in arm B [position D2 in Fig.~\ref{fig:phantom}(a)], slightly tilted down to produce a small vertical displacement of the beam. 

$\hat{\sigma}_A$ and $\hat{\sigma}_B$ could be implemented with a birefringent crystal producing a small polarisation-dependent horizontal beam displacement~\cite{dressel14a}. The eigenstates of  $\hat{\sigma}_A$ and $\hat{\sigma}_B$ are left-elliptical $\ket{L}$ and right-elliptical $\ket{R}$ states; so for these polarisations, the refractive properties of the birefringent crystal should be so that the beam deflects left and right respectively. For other polarisation, a linear superposition of these basis states, the birefringent crystal would deflect left and right with polarisation dependent probability. 

For weak measurements, the system state should be minimally disturbed; this means that the deflections should be less than the characteristic cross-section width of the beam, so that it is uncertain whether an individual photon has been deflected of not. Because of this, the experiment needs run to over a large ensemble to get the average of a single property.

As a specific implementation example, let us consider when the interferometer beam is a Gaussian so that
\begin{equation}
	\Exp{x|m_i}=\Big(\frac{1}{2\pi\sigma^2}\Big)^{1/4}\exp\Big(-\frac{x^2}{4\sigma^2}\Big)~.
\end{equation}

For real WVs, from Eq.~(\ref{eq:mf}) the final pointer state projected onto the position basis is
\begin{equation}
	\Exp{x|m_f}\approx\frac{e^{i\theta}}{2}\Big(\frac{1}{2\pi\sigma^2}\Big)^{1/4}\exp\Big[-\frac{(x-g\Exp{\hat{O}}_w)^2}{4\sigma^2}\Big]~.
\end{equation}

In other words, the Gaussian beam maintains its profile, but the interaction with the projectors deflects it by an amount proportional to the WV. 

For $\Exp{\hat{\sigma}_A}_w = 0$  and $\Exp{\hat{\sigma}_B}_w = i$ we see no deflection in the beam on average; however there will be a difference in the total probability of detection between the two cases. In the position basis
\begin{equation}
	\Exp{\hat{P}}_w = \frac{\bra{x}\hat{P}\ket{m_i}}{\langle x |m_i\rangle } = -i\hbar\frac{\partial_x m_i(x)}{m_i(x)}=i\hbar\frac{x}{2\sigma^2}~.
\end{equation}
This means that $\text{Re}\Exp{\hat{P}}_w=0$, and from Eq.~(\ref{eq:PRatio}) we do not observe any disturbance in the probability of detection for $\Exp{\hat{\sigma}_A}_w = 0$  and $\Exp{\hat{\sigma}_B}_w = i$. However if we measure in the momentum basis, then
\begin{equation}
	\Exp{\hat{P}}_w = \frac{\bra{P}\hat{P}\ket{m_i}}{\langle P |m_i\rangle } = \frac{pm_i(p)}{m_i(p)}=p~.
\label{eq:Pmom}
\end{equation}
This can be implemented with a Fourier lens, so that each pixel on the CCD corresponds to a transverse momentum~(Fig.~\ref{fig:phantom}). Using Eq.~(\ref{eq:Pmom}) in Eq.~(\ref{eq:PRatio}), we see no disturbance in the probability of detection for $\Exp{\hat{\sigma}_A}_w = 0$, but a disturbance in the probability of detection proportional to the transverse momentum of the Gaussian beam for $\Exp{\hat{\sigma}_B}_w = i$, thereby further distinguishing between no polarisation and polarisation.

Two actual attempts of realising the QCC have been conducted with neutron~\cite{denkmayr14} and photonic~\cite{ashby16} interferometry. Although these experiments claimed to have produced the results of the \textit{gedanken} experiment, they have been criticised for not actually implementing the QCC, as they do not make weak measurements~\cite{duprey17}.

\section{Dual of the Quantum Cheshire Cat}
\label{sec:Slit}

Let us consider what would happen if the role of polarisation and location were reversed in the QCC, \textit{i.e.} under the transformation
\begin{equation}
	\ket{A} \leftrightarrow \ket{H}~,\quad \ket{B} \leftrightarrow \ket{V}~. 
\label{eq:map}
\end{equation}

For consistency with Sec.~\ref{sec:Quantum Cheshire Cat}, we will also use $\theta$ to indicate the phase difference between the arms and $\phi$ the polarization phase. 
The post-selection state is invariant under this transformation, but the pre-selection state changes to
\begin{equation}
	\ket{\Psi_i} = (e^{i\phi}\ket{H} + \ket{V})\ket{A}/\sqrt{2}~.
\label{eq:PS_in}
\end{equation}

The projectors corresponding to the transformation are
\begin{align}
	\hat{H}&=\ket{H}\bra{H} (\ket{+}\bra{+} + \ket{-}\bra{-})~,\label{eq:H}\\
	\hat{V}&=\ket{V}\bra{V} (\ket{+}\bra{+} + \ket{-}\bra{-})~,\label{eq:V}
\end{align}
and
\begin{align}
	\hat{\sigma}_H &= \ket{H}\bra{H}(\ket{+}\bra{+} - \ket{-}\bra{-})~,\label{eq:sigmaL}\\
	\hat{\sigma}_V &= \ket{V}\bra{V}(\ket{+}\bra{+} - \ket{-}\bra{-})~,\label{eq:sigma}
\end{align}
where
\begin{align}
\ket{+}\equiv (\ket{A}+e^{i\theta}\ket{B})/\sqrt{2}~,\label{eq:+}\\
\ket{-}\equiv (\ket{A}-e^{i\theta}\ket{B})/\sqrt{2}~.\label{eq:-}
\end{align}

$\hat{H}$ and $\hat{V}$ detect whether the photon is horizontally or vertically polarised. $\hat{\sigma}_H$ detects the phase difference between the arms, for horizontal polarisation.  The eigenvalues 1 and -1 correspond to eigenstates $\ket{+,H}$ and $\ket{-,H}$ respectively; whereas the eigenvalue 0 corresponds to the degenerate subspace spanned by eigenstates $\ket{+,V}$ and $\ket{-,V}$. $\hat{\sigma}_V$ is similarly defined.  

Using Eq.~(\ref{eq:weak}), the WVs measured are
\begin{align}
	\Exp{\hat{H}}_w &=\Exp{\hat{H}\hat{p}}_w + \Exp{\hat{H}\hat{m}}_w= 1~,\label{eq:hatH}\\
	\Exp{\hat{V}}_w &=\Exp{\hat{V}\hat{p}}_w + \Exp{\hat{V}\hat{m}}_w= 0~\label{eq:hatV}
\end{align} 
and
\begin{align}
	\Exp{\hat{\sigma}_H}_w &=\Exp{\hat{H}\hat{p}}_w - \Exp{\hat{H}\hat{m}}_w= 0~,\label{eq:hatSigmaH}\\
	\Exp{\hat{\sigma}_V}_w &= \Exp{\hat{V}\hat{p}}_w - \Exp{\hat{V}\hat{m}}_w=e^{i(\phi-\theta)}~,\label{eq:hatSigmaV}
\end{align} 
where
\begin{align}
	\hat{p}&=\ket{+}\bra{+} (\ket{H}\bra{H} + \ket{V}\bra{V})~,\\
	\hat{m}&=\ket{-}\bra{-} (\ket{H}\bra{H} + \ket{V}\bra{V})~.
\end{align} 
Note that $\Exp{\hat{H}\hat{p}}_w=\Exp{\hat{H}\hat{m}}_w=1/2$ and $\Exp{\hat{V}\hat{p}}_w=-\Exp{\hat{V}\hat{m}}_w=e^{i(\phi-\theta)}/2$. In this dual to the QCC, the photons are detected to be horizontally polarised, but the phase difference between the two arms is vertically polarised. What is even more remarkable is that in the pre-selection state, there is no arm B. In other words, it is as if the detector is detecting interference between arm A and a phantom arm B! The detected phase difference between arm A and B is determined by the photon's polarisation in the pre-selected state. Specifically, by rotating the basis of the phase operators so that $\theta=\phi$ in Eq.~(\ref{eq:+}) and (\ref{eq:-}), one gets $\Exp{\hat{\sigma}_V}_w =1$. Analogous to the QCC, this is interpreted to mean that we have detected a phase difference, in the vertically-polarised component, between arm A and arm B.

Rewriting
\begin{equation}
	\hat{\sigma}_V =\ket{V}\bra{V}(e^{-i\theta}\ket{A}\bra{B} + e^{i\theta}\ket{B}\bra{A})~,
\label{eq:sigmaV_2}
\end{equation}
we see that $\hat{\sigma}_V$ has the property of flipping $\ket{A}$ and $\ket{B}$ (similarly for $\hat{\sigma}_H$). It is this property that underlies the apparent paradoxical phantom arm effect. This is exactly analogous to the central role that 
\begin{equation}
	\hat{\sigma}_B =\ket{B}\bra{B}(e^{-i\phi}\ket{H}\bra{V} + e^{i\phi}\ket{V}\bra{H})
\label{eq:sigmaB_2}
\end{equation}
(and similarly for $\hat{\sigma}_A$) plays in the QCC.

\subsection{Implementation}

Underpinning the QCC were the circular-polarisation $\hat{\sigma}_A$ and $\hat{\sigma}_B$ detectors. They were proposed to be implemented as optical elements which slightly deflects the beam left (right) for left- (right-) circular polarisation, from the axis of arm A or B. This slight deflection provides the means to read the weak values on the CCD camera. Upon interaction, the detectors turns horizontal polarisation vertical and vice versa, as described by Eq.~(\ref{eq:sigmaB_2}). However, for weak interactions, this only occurs for a small subset of photons.  Likewise, $\hat{\sigma}_H$ and $\hat{\sigma}_V$ should cause a slight left (right) deflection when the beams are in (out-of) phase. For a small subset of photon, arm A turns into arm B, as described by Eq.~(\ref{eq:sigmaV_2}). This can be achieved by placing a PBS  at position D1 in Fig.~\ref{fig:phantom_camera}. The unitary transformation for the $\hat{\sigma}_V$ detector (similarly for $\hat{\sigma}_H$) is~\cite{campos89}
\begin{equation}
	\hat{U}=e^{\gamma\hat{\sigma}_V}~.
\end{equation}
The PBS is transparent to horizontal polarisation, but partially reflects vertical polarisation, with reflectivity $r_V=\sin^2\gamma$ and transmissivity $t_V=1-r_V$. 

Considering just the vertical polarisation component, we write $\ket{A}$ and $\ket{B}$ in terms of creation operators on the vacuum state $\ket{0}$,
\begin{equation}
	\ket{A}=\hat{a}^\dagger_{A}\ket{0}~,\quad \ket{B}=\hat{a}^\dagger_{B}\ket{0}~.
\end{equation}
In the Heisenberg picture, these operators transform under the PBSs as
\begin{equation}
	\hat{U}^\dagger
		\begin{pmatrix}
	 		\hat{a}_{A} \\
	 		\hat{a}_{B}
		\end{pmatrix}
	\hat{U}\equiv
		\begin{pmatrix}
		\hat{a}_{A'} \\
		\hat{a}_{B'}
		\end{pmatrix}~.
\end{equation}
Using the Baker-Hausdorf lemma,
\begin{align}
	\hat{a}_{A'}&=\hat{a}_{A}\cos\gamma+\hat{a}_{B}e^{-i\theta}\sin\gamma~,\label{eq:aA'}\\
	\hat{a}_{B'}&=-\hat{a}_{A}e^{i\theta}\sin\gamma+\hat{a}_{B}\cos\gamma~.\label{eq:aB'}
\end{align}
From Eq.~(\ref{eq:aA'}) and Eq.~(\ref{eq:aB'}) we get,
%\begin{equation}
%\begin{split}
%	\ket{A}&=\hat{a}_A\ket{0}\\ 
%		&= (\hat{a}_{A'}\cos\gamma-\hat{a}_{B'}e^{-i\theta}\sin\gamma)\ket{0}\\
%		&= \cos\gamma\ket{A'}-e^{-i\theta}\sin\gamma\ket{B'}
%\end{split}
%\end{equation}
\begin{gather}
	\hat{a}_A = \hat{a}_{A'}\cos\gamma-\hat{a}_{B'}e^{-i\theta}\sin\gamma~,\label{eq:aA}\\
	%\hat{a}_B\ket{0} = \hat{a}_{A'}e^{-i\theta}\sin\gamma+\hat{a}_{B'}\cos\gamma~,
	\hat{a}_A\pm e^{-i\theta}\hat{a}_B= \hat{a}_{A'}(\cos\gamma\pm\sin\gamma)\pm\hat{a}_{B'}e^{-i\theta}(\cos\gamma\mp\sin\gamma)~.\label{eq:aApmaB}
\end{gather}
Applying the conjugates of the operators in Eq.~({\ref{eq:aA}) and (\ref{eq:aApmaB}) to the vacuum state, we see that the PBS transforms
\begin{gather}
	\ket{A}\rightarrow \cos\gamma\ket{A'}-e^{-i\theta}\sin\gamma\ket{B'}~,\label{eq:Atransform}\\
	\ket{\pm}\rightarrow [(\cos\gamma\pm\sin\gamma)\ket{A'}\pm e^{-i\theta}(\cos\gamma\mp\sin\gamma)\ket{B'}]/\sqrt{2}~\label{eq:pmTransform}.
\end{gather}
Eq.~(\ref{eq:aA}) - (\ref{eq:pmTransform}) equally applies to the horizontal-polarisation component for $\hat{\sigma}_H$.
	
Fig.~\ref{fig:BS} illustrates Eq.~(\ref{eq:Atransform}) and (\ref{eq:pmTransform}) for $\gamma=\pi/4$.  The eigenstates of $\hat{\sigma}_H$ and $\hat{\sigma}_V$ are the $\ket{+}$ and $\ket{-}$ states; so for such states the detector deflects the beam left and right respectively. For states composing of just one arm, the detector deflects left and right. The deflection should be slightly off the axis of the beam A or B; as such the PBS should be slightly wedged. $\theta=2\pi d/ \lambda$ is the phase difference between the reflected and transmitted beams; it is determined by the path difference $d$ due to the width of the PBS, and wavelength $\lambda$. For weak interactions, $\gamma$ should be small, so that only a small subset of photons are deflected. Furthermore, the deflection $\delta$ should not be greater than the cross-sectional width of the beam. 

\begin{figure}
	\centering
	\includegraphics[width=\columnwidth]{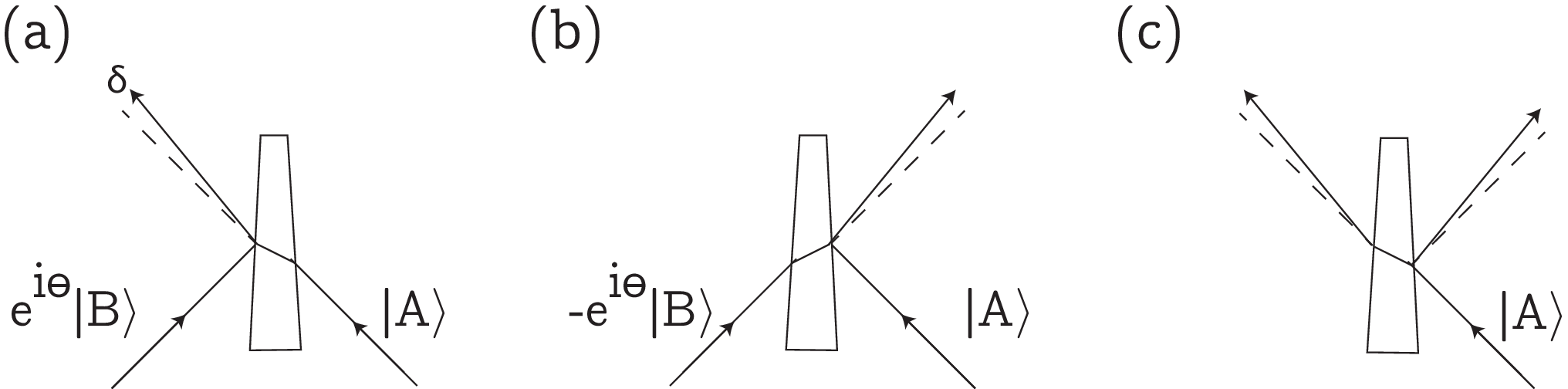}
	\caption{$\hat{\sigma}_H$ ($\hat{\sigma}_V$) could be implemented with a wedged PBS. This figure illustrates Eq.~(\ref{eq:Atransform}) and (\ref{eq:pmTransform}) for $\gamma=\pi/4$.  The eigenstates of $\hat{\sigma}_H$ and $\hat{\sigma}_V$ are the $\ket{+}$ and $\ket{-}$ states; for such states the detector deflects the beam (a) left and (b) right respectively. (c) For states composing of just one arm, the detector deflects left and right. The PBS is wedged so the deflection is slightly off the axis of the beam A or B (dotted lines). This small deflection $\delta$ provides the means to read the weak values on the CCD camera. For weak measurements, $\delta$ should not be greater than the cross-sectional width of the beam. }
	\label{fig:BS}
\end{figure}

The PBS can simply implemented as a piece of glass. The Fresnel reflection are different for the polarisation component in the plane of incidence and the orthogonal component. We can define coordinates so that horizontal-polarisation is in the plane of incidence, so that the reflectivity are
\begin{equation}
	r_H=\Big[\frac{\tan(\nu_i-\nu_t)}{\tan(\nu_i+\nu_t)}\Big]^2~,\quad r_V=\Big[\frac{\sin(\nu_i-\nu_t)}{\sin(\nu_i+\nu_t)}\Big]^2~,
\end{equation}
where $\nu_i$ ($\nu_t$) is the angle of incidence (transmission). 

$r_H=0$ at the Brewster angle  $\nu_i=\arctan(n)$, where $n\approx1.5$ is the refractive index of glass. Fig.~\ref{fig:reflection} plots reflectivity as function of $n$, for $\nu_i=\arctan(1.5)\approx56^\circ$. At the Brewster angle $r_V\approx0.15$. This satisfies the requirement that $\hat{\sigma}_V$ transmits horizontal polarisation, but reflects vertical polarisation with small probability. For $\hat{\sigma}_H$ one would orient the piece of glass (and experiment) so that the vertical-polarisation component is in the plane of incidence. 

\begin{figure}
	\centering
	\includegraphics[width=0.8\columnwidth]{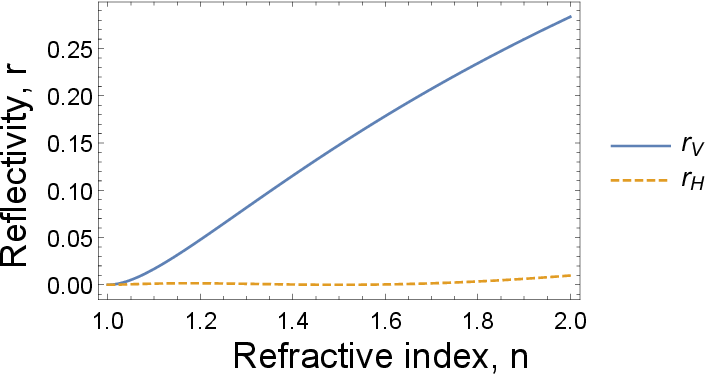}
	\caption{A plot of reflectivity $r_V$ and $r_H$ for incident angle $\nu_i=\arctan(1.5)\approx56^\circ$, for refractive indices between 1 and 2. For $n=1.5$, $\nu_i=\arctan(1.5)$ is the Brewster angle with $r_H=0$ and $r_V\approx0.15$. This satisfies the requirement that $\hat{\sigma}_V$ transmits horizontal polarisation, but reflects vertical polarisation with small probability. }
	\label{fig:reflection}
\end{figure}

\begin{figure}
	\centering
	\includegraphics[width=0.5\columnwidth]{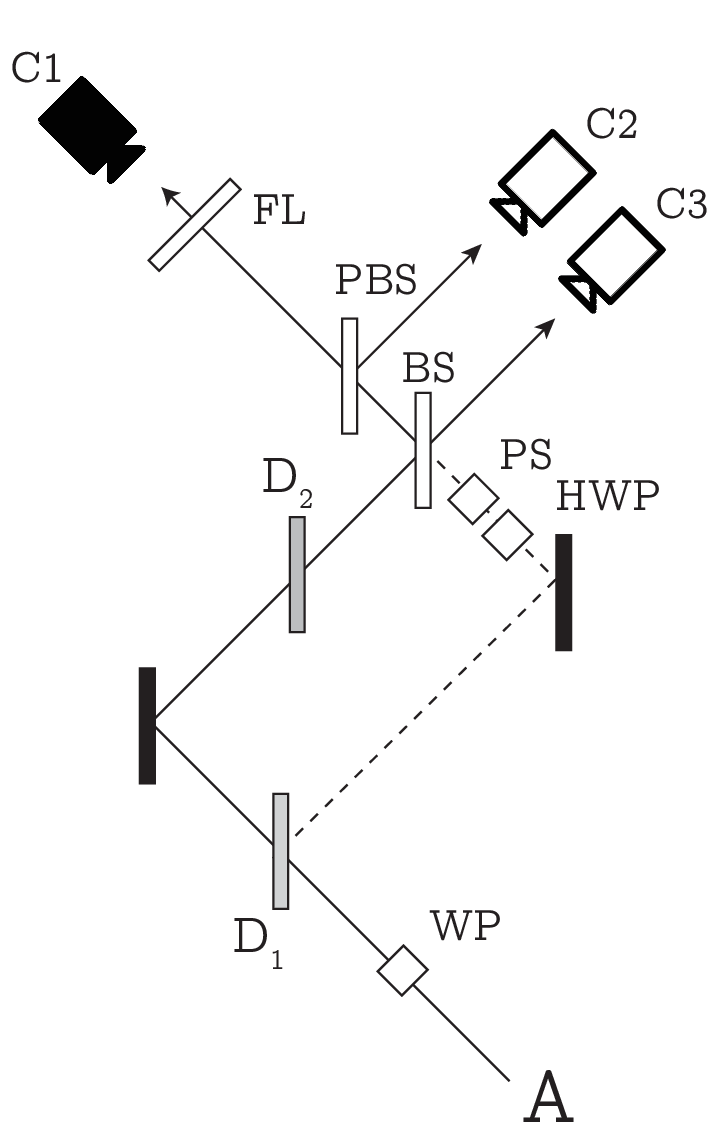}
	\caption{A schematic of the phantom arm, where a wave-plate (WP) introduces a controllable phase $e^{i\phi}$.  A half-wave plate (HWP), phase-shifter (PS), beam-splitter (BS), and polarising beam-splitter (PBS), are used for post-selection. To project the pointer onto the momentum basis, a Fourier lens (FL) is used to Fourier transform the light beam so that each pixel on the camera corresponds to a transverse momentum. To project the pointer onto the position basis, remove the FL, and each pixel corresponds to a transverse displacement.  Detectors $\hat{\sigma}_H$ ($\hat{\sigma}_V$) are placed at location $D_1$, and $\hat{H}$ ($\hat{V}$) at location $D_2$. The solid line represents the  axes of the main beam, and the dashed line represents the axes of the weak beam after interaction with the detector at position $D_1$.}
\label{fig:phantom_camera}
	\label{fig:phantom_camera}
\end{figure}

$\hat{H}$ and $\hat{V}$ could be a birefringent crystal producing a small polarisation-dependent horizontal beam displacement, placed at position D2 in Fig.~\ref{fig:phantom_camera}. As with the QCC, the detectors only minimally disturb the beam and therefore can be measured simultaneously. In particular, measuring $\Exp{\hat{\sigma}_V}_w$ and$\Exp{\hat{H}}_w$ simultaneously, will show that the phase difference is vertically polarised, but the photon is horizontally polarised. Note that to measure $\hat{\sigma}_V$ and $\hat{\sigma}_H$ simultaneously, one needs to duplicate the setup depicted in Fig.~\ref{fig:phantom_camera}, for the two beams reflected by the detectors.

\subsection{Temporal Interference} 
\label{sec:double-slit experiment}

At the heart of the QCC and its dual is the interplay between the pre-selected and post-selected states; this give rise to the concept of temporal interference in the QCC dual. To understand this idea, we compare it with a conventional notion of interference. In the double-slit experiment, every point on the detection screen can be considered as the interference between two light rays emanating from the slits. As rays from the two slits must travel different distances to different points on the screen, a phase differential arises. If we label the state of a photon emanating from slit $\alpha$ ($\beta$) as $\ket{\alpha}$ ($\ket{\beta}$), the interference is the result of the phase-differential that arises from the spatial separation of $\ket{\alpha}$ and $\ket{\beta}$. $\ket{\alpha}$ is spatially separated from $\ket{\beta}$ in the sense that one is the spatial translation of the other. 

The spatial separation of the slits is implemented in an interferometer with a controllable phase-shift in one of the arms, as in $\ket{\Phi_i}$, where $\ket{A}$ and $\ket{B}$ would represent states of photons that went through slit $\alpha$  and $\beta$. In comparison, in $\ket{\Psi_i}$ there are no spatially separated states as there is just one arm of the interferometer (or equivalently just one slit). In this case, the detected interference pattern arises from the phase differential between the pre-selected and post-selected states; in other words, it is a temporal interference between past and future quantum states.

Reinforcing this notion, the pre-selection and post-selection states themselves do not exhibit interference ($\hat{\sigma}=\hat{\sigma}_H \text{ or } \hat{\sigma}_V$): $\Exp{\Psi_i|\hat{\sigma}|\Psi_i} = \Exp{\Psi_f|\hat{\sigma}|\Psi_f} = 0$~. Whereas, in comparison  $\Exp{\Psi_f|\hat{\sigma}|\Psi_i}$ is in general non-zero. This is analogous to the fact that the $\ket{\alpha}$ and $\ket{\beta}$ do not exhibit interference on their own but rather it arises from the phase difference between them. 
%$\Exp{\Psi_i|\hat{\sigma}_H|\Psi_i} = \Exp{\Psi_i|\hat{\sigma}_V|\Psi_i} = \Exp{\Psi_f|\hat{\sigma}_H|\Psi_f} = \Exp{\Psi_f|\hat{\sigma}_V|\Psi_f} = 0$

The notion of WV temporal interference has been demonstrated with a driven superconducting qubit~\cite{campagne14}. In this experiment the fluorescence of a qubit prepared in the ground and excited states were each measured. Then the fluorescence of the pre-selected excited state, post-selected to be in the ground state was measured, and it showed a pattern which appeared to be the interference between the fluorescence of the previously measured ground and excited state patterns. The effect we propose here goes beyond this, as we propose that the interference is carried by horizontally-polarised photons, yet the photons are detected to be vertically polarised. 

We also point out that temporal uncertainty can give rise to frequency domain interference, in systems with time-dependent amplitude modulation~\cite{hauser74}. The dual of the QCC is distinctly different from this type of temporal interference, as the interference we describe is in the position domain and is the result of interference between past and future quantum states.

\section{Outlook}
\label{sec:Conclusion}
By generalising the QCC with elliptical polarisation, we showed that the WV of the polarisation for the pre- and post-selected states of the generalised QCC, is determined by the phase difference in the interferometer. We also showed that the generalisation provides a consistent way to interpret the zero WV of this polarisation.

We explored the novel behaviour of the position-polarisation duality in the QCC. We have shown in the QCC dual that whilst the photons are horizontally polarised, their phases are vertically polarised. The QCC dual gives rise to temporal interference. The past and future states are particle-like, whilst the intermediate states are wave-like in that they exhibit interference. 

In an experiment that directly addresses the question of the physical reality of observables before wave function collapse, Kocsis \textit{et. al}~\cite{kocsis11} used WVs to observe the individual trajectories of photons as they formed the interference pattern in the double-slit experiment. More recently, individual quantum trajectories of a superconducting circuit were also reconstructed using weak measurements~\cite{weber14}. As the interferometer setup in this work can map to the double-slit experiment, it would be interesting in future work to map the trajectories of individual photons, along the lines of the Kocsis \textit{et. al} experiment, to see how the interference pattern arises as the photon evolves from the pre-selected to post-selected state in the QCC dual. In addition, as the QCC has not been satisfactorily realised in experiments, the dual QCC setup offers an alternative pathway to realisation.

\section*{Acknowledgements}
\label{sec:Acknowledgements}
The author thanks  M. Lewenstein, M. Bera, A. Riera, and S. Quach for discussions and checking the manuscript. This work was financially supported by the Ramsay fellowship.

% the European Commission's Marie Curie Actions for Co-funding of Regional, National and International Programmes (COFUND), the Spanish Ministry MINECO (National Plan Grant: FISICATEAMO No. FIS2016-79508-P, SEVERO OCHOA No. SEV-2015-0522), Fundació Privada Cellex, Generalitat de Catalunya (AGAUR Grant No. 2014 SGR874 and CERCA/Program), ERC AdG OSYRIS, and EU FETPRO QUIC. 

%\bibliographystyle{plainnat}
\bibliography{cheshire}

\begin{thebibliography}{50}%
\makeatletter
\providecommand \@ifxundefined [1]{%
 \@ifx{#1\undefined}
}%
\providecommand \@ifnum [1]{%
 \ifnum #1\expandafter \@firstoftwo
 \else \expandafter \@secondoftwo
 \fi
}%
\providecommand \@ifx [1]{%
 \ifx #1\expandafter \@firstoftwo
 \else \expandafter \@secondoftwo
 \fi
}%
\providecommand \natexlab [1]{#1}%
\providecommand \enquote  [1]{``#1''}%
\providecommand \bibnamefont  [1]{#1}%
\providecommand \bibfnamefont [1]{#1}%
\providecommand \citenamefont [1]{#1}%
\providecommand \href@noop [0]{\@secondoftwo}%
\providecommand \href [0]{\begingroup \@sanitize@url \@href}%
\providecommand \@href[1]{\@@startlink{#1}\@@href}%
\providecommand \@@href[1]{\endgroup#1\@@endlink}%
\providecommand \@sanitize@url [0]{\catcode `\\12\catcode `\$12\catcode
  `\&12\catcode `\#12\catcode `\^12\catcode `\_12\catcode `\%12\relax}%
\providecommand \@@startlink[1]{}%
\providecommand \@@endlink[0]{}%
\providecommand \url  [0]{\begingroup\@sanitize@url \@url }%
\providecommand \@url [1]{\endgroup\@href {#1}{\urlprefix }}%
\providecommand \urlprefix  [0]{URL }%
\providecommand \Eprint [0]{\href }%
\providecommand \doibase [0]{http://dx.doi.org/}%
\providecommand \selectlanguage [0]{\@gobble}%
\providecommand \bibinfo  [0]{\@secondoftwo}%
\providecommand \bibfield  [0]{\@secondoftwo}%
\providecommand \translation [1]{[#1]}%
\providecommand \BibitemOpen [0]{}%
\providecommand \bibitemStop [0]{}%
\providecommand \bibitemNoStop [0]{.\EOS\space}%
\providecommand \EOS [0]{\spacefactor3000\relax}%
\providecommand \BibitemShut  [1]{\csname bibitem#1\endcsname}%
\let\auto@bib@innerbib\@empty
%</preamble>
\bibitem [{\citenamefont {Aharonov}\ \emph {et~al.}(1988)\citenamefont
  {Aharonov}, \citenamefont {Albert},\ and\ \citenamefont
  {Vaidman}}]{aharonov88}%
  \BibitemOpen
  \bibfield  {author} {\bibinfo {author} {\bibfnamefont {Y.}~\bibnamefont
  {Aharonov}}, \bibinfo {author} {\bibfnamefont {D.~Z.}\ \bibnamefont
  {Albert}}, \ and\ \bibinfo {author} {\bibfnamefont {L.}~\bibnamefont
  {Vaidman}},\ }\bibfield  {title} {\enquote {\bibinfo {title} {How the result
  of a measurement of a component of the spin of a spin-1/2 particle can turn
  out to be 100},}\ }\href {\doibase 10.1103/PhysRevLett.60.1351} {\bibfield
  {journal} {\bibinfo  {journal} {Phys. Rev. Lett.}\ }\textbf {\bibinfo
  {volume} {60}},\ \bibinfo {pages} {1351--1354} (\bibinfo {year}
  {1988})}\BibitemShut {NoStop}%
\bibitem [{\citenamefont {Aharonov}\ and\ \citenamefont
  {Vaidman}(2002)}]{aharonov02}%
  \BibitemOpen
  \bibfield  {author} {\bibinfo {author} {\bibfnamefont {Y.}~\bibnamefont
  {Aharonov}}\ and\ \bibinfo {author} {\bibfnamefont {L.}~\bibnamefont
  {Vaidman}},\ }\enquote {\bibinfo {title} {The two-state vector formalism of
  quantum mechanics},}\ in\ \href {\doibase 10.1007/3-540-45846-8_13} {\emph
  {\bibinfo {booktitle} {Time in Quantum Mechanics}}},\ \bibinfo {editor}
  {edited by\ \bibinfo {editor} {\bibfnamefont {J.~G.}\ \bibnamefont {Muga}},
  \bibinfo {editor} {\bibfnamefont {R.~Sala}\ \bibnamefont {Mayato}}, \ and\
  \bibinfo {editor} {\bibfnamefont {I.~L.}\ \bibnamefont {Egusquiza}}}\
  (\bibinfo  {publisher} {Springer Berlin Heidelberg},\ \bibinfo {address}
  {Berlin, Heidelberg},\ \bibinfo {year} {2002})\ pp.\ \bibinfo {pages}
  {369--412}\BibitemShut {NoStop}%
\bibitem [{\citenamefont {Brunner}\ and\ \citenamefont
  {Simon}(2010)}]{brunner10}%
  \BibitemOpen
  \bibfield  {author} {\bibinfo {author} {\bibfnamefont {N.}~\bibnamefont
  {Brunner}}\ and\ \bibinfo {author} {\bibfnamefont {C.}~\bibnamefont
  {Simon}},\ }\bibfield  {title} {\enquote {\bibinfo {title} {Measuring small
  longitudinal phase shifts: Weak measurements or standard interferometry?}}\
  }\href {\doibase 10.1103/PhysRevLett.105.010405} {\bibfield  {journal}
  {\bibinfo  {journal} {Phys. Rev. Lett.}\ }\textbf {\bibinfo {volume} {105}},\
  \bibinfo {pages} {010405} (\bibinfo {year} {2010})}\BibitemShut {NoStop}%
\bibitem [{\citenamefont {Dixon}\ \emph {et~al.}(2009)\citenamefont {Dixon},
  \citenamefont {Starling}, \citenamefont {Jordan},\ and\ \citenamefont
  {Howell}}]{dixon09}%
  \BibitemOpen
  \bibfield  {author} {\bibinfo {author} {\bibfnamefont {P.~B.}\ \bibnamefont
  {Dixon}}, \bibinfo {author} {\bibfnamefont {D.~J.}\ \bibnamefont {Starling}},
  \bibinfo {author} {\bibfnamefont {A.~N.}\ \bibnamefont {Jordan}}, \ and\
  \bibinfo {author} {\bibfnamefont {J.~C.}\ \bibnamefont {Howell}},\ }\bibfield
   {title} {\enquote {\bibinfo {title} {Ultrasensitive beam deflection
  measurement via interferometric weak value amplification},}\ }\href {\doibase
  10.1103/PhysRevLett.102.173601} {\bibfield  {journal} {\bibinfo  {journal}
  {Phys. Rev. Lett.}\ }\textbf {\bibinfo {volume} {102}},\ \bibinfo {pages}
  {173601} (\bibinfo {year} {2009})}\BibitemShut {NoStop}%
\bibitem [{\citenamefont {Egan}\ and\ \citenamefont {Stone}(2012)}]{egan12}%
  \BibitemOpen
  \bibfield  {author} {\bibinfo {author} {\bibfnamefont {P.}~\bibnamefont
  {Egan}}\ and\ \bibinfo {author} {\bibfnamefont {J.~A.}\ \bibnamefont
  {Stone}},\ }\bibfield  {title} {\enquote {\bibinfo {title} {Weak-value
  thermostat with 0.2 mk precision},}\ }\href {\doibase 10.1364/OL.37.004991}
  {\bibfield  {journal} {\bibinfo  {journal} {Opt. Lett.}\ }\textbf {\bibinfo
  {volume} {37}},\ \bibinfo {pages} {4991--4993} (\bibinfo {year}
  {2012})}\BibitemShut {NoStop}%
\bibitem [{\citenamefont {Feizpour}\ \emph {et~al.}(2011)\citenamefont
  {Feizpour}, \citenamefont {Xing},\ and\ \citenamefont
  {Steinberg}}]{feizpour11}%
  \BibitemOpen
  \bibfield  {author} {\bibinfo {author} {\bibfnamefont {A.}~\bibnamefont
  {Feizpour}}, \bibinfo {author} {\bibfnamefont {X.}~\bibnamefont {Xing}}, \
  and\ \bibinfo {author} {\bibfnamefont {A.~M.}\ \bibnamefont {Steinberg}},\
  }\bibfield  {title} {\enquote {\bibinfo {title} {Amplifying single-photon
  nonlinearity using weak measurements},}\ }\href {\doibase
  10.1103/PhysRevLett.107.133603} {\bibfield  {journal} {\bibinfo  {journal}
  {Phys. Rev. Lett.}\ }\textbf {\bibinfo {volume} {107}},\ \bibinfo {pages}
  {133603} (\bibinfo {year} {2011})}\BibitemShut {NoStop}%
\bibitem [{\citenamefont {Hogan}\ \emph {et~al.}(2011)\citenamefont {Hogan},
  \citenamefont {Hammer}, \citenamefont {Chiow}, \citenamefont {Dickerson},
  \citenamefont {Johnson}, \citenamefont {Kovachy}, \citenamefont
  {Sugarbaker},\ and\ \citenamefont {Kasevich}}]{hogan11}%
  \BibitemOpen
  \bibfield  {author} {\bibinfo {author} {\bibfnamefont {J.~M.}\ \bibnamefont
  {Hogan}}, \bibinfo {author} {\bibfnamefont {J.}~\bibnamefont {Hammer}},
  \bibinfo {author} {\bibfnamefont {S.-W.}\ \bibnamefont {Chiow}}, \bibinfo
  {author} {\bibfnamefont {S.}~\bibnamefont {Dickerson}}, \bibinfo {author}
  {\bibfnamefont {D.~M.~S.}\ \bibnamefont {Johnson}}, \bibinfo {author}
  {\bibfnamefont {T.}~\bibnamefont {Kovachy}}, \bibinfo {author} {\bibfnamefont
  {A.}~\bibnamefont {Sugarbaker}}, \ and\ \bibinfo {author} {\bibfnamefont
  {M.~A.}\ \bibnamefont {Kasevich}},\ }\bibfield  {title} {\enquote {\bibinfo
  {title} {Precision angle sensor using an optical lever inside a sagnac
  interferometer},}\ }\href {\doibase 10.1364/OL.36.001698} {\bibfield
  {journal} {\bibinfo  {journal} {Opt. Lett.}\ }\textbf {\bibinfo {volume}
  {36}},\ \bibinfo {pages} {1698--1700} (\bibinfo {year} {2011})}\BibitemShut
  {NoStop}%
\bibitem [{\citenamefont {Hosten}\ and\ \citenamefont
  {Kwiat}(2008)}]{hosten08}%
  \BibitemOpen
  \bibfield  {author} {\bibinfo {author} {\bibfnamefont {O.}~\bibnamefont
  {Hosten}}\ and\ \bibinfo {author} {\bibfnamefont {P.}~\bibnamefont {Kwiat}},\
  }\bibfield  {title} {\enquote {\bibinfo {title} {Observation of the spin hall
  effect of light via weak measurements},}\ }\href {\doibase
  10.1126/science.1152697} {\ \textbf {\bibinfo {volume} {319}},\ \bibinfo
  {pages} {787--790} (\bibinfo {year} {2008})}\BibitemShut {NoStop}%
\bibitem [{\citenamefont {Jayaswal}\ \emph {et~al.}(2014)\citenamefont
  {Jayaswal}, \citenamefont {Mistura},\ and\ \citenamefont
  {Merano}}]{jayaswal14}%
  \BibitemOpen
  \bibfield  {author} {\bibinfo {author} {\bibfnamefont {G.}~\bibnamefont
  {Jayaswal}}, \bibinfo {author} {\bibfnamefont {G.}~\bibnamefont {Mistura}}, \
  and\ \bibinfo {author} {\bibfnamefont {M.}~\bibnamefont {Merano}},\
  }\bibfield  {title} {\enquote {\bibinfo {title} {Observation of the
  imbert-fedorov effect via weak value amplification},}\ }\href {\doibase
  10.1364/OL.39.002266} {\bibfield  {journal} {\bibinfo  {journal} {Opt.
  Lett.}\ }\textbf {\bibinfo {volume} {39}},\ \bibinfo {pages} {2266--2269}
  (\bibinfo {year} {2014})}\BibitemShut {NoStop}%
\bibitem [{\citenamefont {Jordan}\ \emph {et~al.}(2014)\citenamefont {Jordan},
  \citenamefont {Mart\'{\i}nez-Rinc\'on},\ and\ \citenamefont
  {Howell}}]{jordan14}%
  \BibitemOpen
  \bibfield  {author} {\bibinfo {author} {\bibfnamefont {A.~N.}\ \bibnamefont
  {Jordan}}, \bibinfo {author} {\bibfnamefont {J.}~\bibnamefont
  {Mart\'{\i}nez-Rinc\'on}}, \ and\ \bibinfo {author} {\bibfnamefont {J.~C.}\
  \bibnamefont {Howell}},\ }\bibfield  {title} {\enquote {\bibinfo {title}
  {Technical advantages for weak-value amplification: When less is more},}\
  }\href {\doibase 10.1103/PhysRevX.4.011031} {\bibfield  {journal} {\bibinfo
  {journal} {Phys. Rev. X}\ }\textbf {\bibinfo {volume} {4}},\ \bibinfo {pages}
  {011031} (\bibinfo {year} {2014})}\BibitemShut {NoStop}%
\bibitem [{\citenamefont {Knee}\ and\ \citenamefont {Gauger}(2014)}]{knee14}%
  \BibitemOpen
  \bibfield  {author} {\bibinfo {author} {\bibfnamefont {G.~C.}\ \bibnamefont
  {Knee}}\ and\ \bibinfo {author} {\bibfnamefont {E.~M.}\ \bibnamefont
  {Gauger}},\ }\bibfield  {title} {\enquote {\bibinfo {title} {When
  amplification with weak values fails to suppress technical noise},}\ }\href
  {\doibase 10.1103/PhysRevX.4.011032} {\bibfield  {journal} {\bibinfo
  {journal} {Phys. Rev. X}\ }\textbf {\bibinfo {volume} {4}},\ \bibinfo {pages}
  {011032} (\bibinfo {year} {2014})}\BibitemShut {NoStop}%
\bibitem [{\citenamefont {Pfeifer}\ and\ \citenamefont
  {Fischer}(2011)}]{pfeifer11}%
  \BibitemOpen
  \bibfield  {author} {\bibinfo {author} {\bibfnamefont {M.}~\bibnamefont
  {Pfeifer}}\ and\ \bibinfo {author} {\bibfnamefont {P.}~\bibnamefont
  {Fischer}},\ }\bibfield  {title} {\enquote {\bibinfo {title} {Weak value
  amplified optical activity measurements},}\ }\href {\doibase
  10.1364/OE.19.016508} {\bibfield  {journal} {\bibinfo  {journal} {Opt.
  Express}\ }\textbf {\bibinfo {volume} {19}},\ \bibinfo {pages} {16508--16517}
  (\bibinfo {year} {2011})}\BibitemShut {NoStop}%
\bibitem [{\citenamefont {Starling}\ \emph {et~al.}(2009)\citenamefont
  {Starling}, \citenamefont {Dixon}, \citenamefont {Jordan},\ and\
  \citenamefont {Howell}}]{starling09}%
  \BibitemOpen
  \bibfield  {author} {\bibinfo {author} {\bibfnamefont {D.~J.}\ \bibnamefont
  {Starling}}, \bibinfo {author} {\bibfnamefont {P.~B.}\ \bibnamefont {Dixon}},
  \bibinfo {author} {\bibfnamefont {A.~N.}\ \bibnamefont {Jordan}}, \ and\
  \bibinfo {author} {\bibfnamefont {J.~C.}\ \bibnamefont {Howell}},\ }\bibfield
   {title} {\enquote {\bibinfo {title} {Optimizing the signal-to-noise ratio of
  a beam-deflection measurement with interferometric weak values},}\ }\href
  {\doibase 10.1103/PhysRevA.80.041803} {\bibfield  {journal} {\bibinfo
  {journal} {Phys. Rev. A}\ }\textbf {\bibinfo {volume} {80}},\ \bibinfo
  {pages} {041803} (\bibinfo {year} {2009})}\BibitemShut {NoStop}%
\bibitem [{\citenamefont {Starling}\ \emph
  {et~al.}(2010{\natexlab{a}})\citenamefont {Starling}, \citenamefont {Dixon},
  \citenamefont {Jordan},\ and\ \citenamefont {Howell}}]{starling10}%
  \BibitemOpen
  \bibfield  {author} {\bibinfo {author} {\bibfnamefont {D.~J.}\ \bibnamefont
  {Starling}}, \bibinfo {author} {\bibfnamefont {P.~B.}\ \bibnamefont {Dixon}},
  \bibinfo {author} {\bibfnamefont {A.~N.}\ \bibnamefont {Jordan}}, \ and\
  \bibinfo {author} {\bibfnamefont {J.~C.}\ \bibnamefont {Howell}},\ }\bibfield
   {title} {\enquote {\bibinfo {title} {Precision frequency measurements with
  interferometric weak values},}\ }\href {\doibase 10.1103/PhysRevA.82.063822}
  {\bibfield  {journal} {\bibinfo  {journal} {Phys. Rev. A}\ }\textbf {\bibinfo
  {volume} {82}},\ \bibinfo {pages} {063822} (\bibinfo {year}
  {2010}{\natexlab{a}})}\BibitemShut {NoStop}%
\bibitem [{\citenamefont {Starling}\ \emph
  {et~al.}(2010{\natexlab{b}})\citenamefont {Starling}, \citenamefont {Dixon},
  \citenamefont {Williams}, \citenamefont {Jordan},\ and\ \citenamefont
  {Howell}}]{starling10a}%
  \BibitemOpen
  \bibfield  {author} {\bibinfo {author} {\bibfnamefont {D.~J.}\ \bibnamefont
  {Starling}}, \bibinfo {author} {\bibfnamefont {P.~B.}\ \bibnamefont {Dixon}},
  \bibinfo {author} {\bibfnamefont {N.~S.}\ \bibnamefont {Williams}}, \bibinfo
  {author} {\bibfnamefont {A.~N.}\ \bibnamefont {Jordan}}, \ and\ \bibinfo
  {author} {\bibfnamefont {J.~C.}\ \bibnamefont {Howell}},\ }\bibfield  {title}
  {\enquote {\bibinfo {title} {Continuous phase amplification with a sagnac
  interferometer},}\ }\href {\doibase 10.1103/PhysRevA.82.011802} {\bibfield
  {journal} {\bibinfo  {journal} {Phys. Rev. A}\ }\textbf {\bibinfo {volume}
  {82}},\ \bibinfo {pages} {011802} (\bibinfo {year}
  {2010}{\natexlab{b}})}\BibitemShut {NoStop}%
\bibitem [{\citenamefont {Str\"ubi}\ and\ \citenamefont
  {Bruder}(2013)}]{strubi13}%
  \BibitemOpen
  \bibfield  {author} {\bibinfo {author} {\bibfnamefont {G.}~\bibnamefont
  {Str\"ubi}}\ and\ \bibinfo {author} {\bibfnamefont {C.}~\bibnamefont
  {Bruder}},\ }\bibfield  {title} {\enquote {\bibinfo {title} {Measuring
  ultrasmall time delays of light by joint weak measurements},}\ }\href
  {\doibase 10.1103/PhysRevLett.110.083605} {\bibfield  {journal} {\bibinfo
  {journal} {Phys. Rev. Lett.}\ }\textbf {\bibinfo {volume} {110}},\ \bibinfo
  {pages} {083605} (\bibinfo {year} {2013})}\BibitemShut {NoStop}%
\bibitem [{\citenamefont {Turner}\ \emph {et~al.}(2011)\citenamefont {Turner},
  \citenamefont {Hagedorn}, \citenamefont {Schlamminger},\ and\ \citenamefont
  {Gundlach}}]{turner11}%
  \BibitemOpen
  \bibfield  {author} {\bibinfo {author} {\bibfnamefont {M.~D.}\ \bibnamefont
  {Turner}}, \bibinfo {author} {\bibfnamefont {C.~A.}\ \bibnamefont
  {Hagedorn}}, \bibinfo {author} {\bibfnamefont {S.}~\bibnamefont
  {Schlamminger}}, \ and\ \bibinfo {author} {\bibfnamefont {J.~H.}\
  \bibnamefont {Gundlach}},\ }\bibfield  {title} {\enquote {\bibinfo {title}
  {Picoradian deflection measurement with an interferometric
  quasi-autocollimator using weak value amplification},}\ }\href {\doibase
  10.1364/OL.36.001479} {\bibfield  {journal} {\bibinfo  {journal} {Opt.
  Lett.}\ }\textbf {\bibinfo {volume} {36}},\ \bibinfo {pages} {1479--1481}
  (\bibinfo {year} {2011})}\BibitemShut {NoStop}%
\bibitem [{\citenamefont {Visa}\ \emph {et~al.}(2013)\citenamefont {Visa},
  \citenamefont {Mart\'{i}nez-Rinc\'{o}n}, \citenamefont {Howland},
  \citenamefont {Frostig}, \citenamefont {Shomroni}, \citenamefont {Dayan},\
  and\ \citenamefont {Howell}}]{viza13}%
  \BibitemOpen
  \bibfield  {author} {\bibinfo {author} {\bibfnamefont {G.~I.}\ \bibnamefont
  {Visa}}, \bibinfo {author} {\bibfnamefont {J.}~\bibnamefont
  {Mart\'{i}nez-Rinc\'{o}n}}, \bibinfo {author} {\bibfnamefont {G.~A.}\
  \bibnamefont {Howland}}, \bibinfo {author} {\bibfnamefont {H.}~\bibnamefont
  {Frostig}}, \bibinfo {author} {\bibfnamefont {I.}~\bibnamefont {Shomroni}},
  \bibinfo {author} {\bibfnamefont {B.}~\bibnamefont {Dayan}}, \ and\ \bibinfo
  {author} {\bibfnamefont {J.~C.}\ \bibnamefont {Howell}},\ }\bibfield  {title}
  {\enquote {\bibinfo {title} {Weak-values technique for velocity
  measurements},}\ }\href {\doibase 10.1364/OL.38.002949} {\bibfield  {journal}
  {\bibinfo  {journal} {Opt. Lett.}\ }\textbf {\bibinfo {volume} {38}},\
  \bibinfo {pages} {2949--2952} (\bibinfo {year} {2013})}\BibitemShut {NoStop}%
\bibitem [{\citenamefont {Zhou}\ \emph {et~al.}(2012)\citenamefont {Zhou},
  \citenamefont {Xiao}, \citenamefont {Luo},\ and\ \citenamefont
  {Wen}}]{zhou12}%
  \BibitemOpen
  \bibfield  {author} {\bibinfo {author} {\bibfnamefont {X.}~\bibnamefont
  {Zhou}}, \bibinfo {author} {\bibfnamefont {Z.}~\bibnamefont {Xiao}}, \bibinfo
  {author} {\bibfnamefont {H.}~\bibnamefont {Luo}}, \ and\ \bibinfo {author}
  {\bibfnamefont {S.}~\bibnamefont {Wen}},\ }\bibfield  {title} {\enquote
  {\bibinfo {title} {Experimental observation of the spin hall effect of light
  on a nanometal film via weak measurements},}\ }\href {\doibase
  10.1103/PhysRevA.85.043809} {\bibfield  {journal} {\bibinfo  {journal} {Phys.
  Rev. A}\ }\textbf {\bibinfo {volume} {85}},\ \bibinfo {pages} {043809}
  (\bibinfo {year} {2012})}\BibitemShut {NoStop}%
\bibitem [{\citenamefont {Zhou}\ \emph {et~al.}(2013)\citenamefont {Zhou},
  \citenamefont {Turek}, \citenamefont {Sun},\ and\ \citenamefont
  {Nori}}]{zhou13}%
  \BibitemOpen
  \bibfield  {author} {\bibinfo {author} {\bibfnamefont {L.}~\bibnamefont
  {Zhou}}, \bibinfo {author} {\bibfnamefont {Y.}~\bibnamefont {Turek}},
  \bibinfo {author} {\bibfnamefont {C.~P.}\ \bibnamefont {Sun}}, \ and\
  \bibinfo {author} {\bibfnamefont {F.}~\bibnamefont {Nori}},\ }\bibfield
  {title} {\enquote {\bibinfo {title} {Weak-value amplification of light
  deflection by a dark atomic ensemble},}\ }\href {\doibase
  10.1103/PhysRevA.88.053815} {\bibfield  {journal} {\bibinfo  {journal} {Phys.
  Rev. A}\ }\textbf {\bibinfo {volume} {88}},\ \bibinfo {pages} {053815}
  (\bibinfo {year} {2013})}\BibitemShut {NoStop}%
\bibitem [{\citenamefont {Kobayashi}\ \emph {et~al.}(2010)\citenamefont
  {Kobayashi}, \citenamefont {Tamate}, \citenamefont {Nakanishi}, \citenamefont
  {Sugiyama},\ and\ \citenamefont {Kitano}}]{kobayashi10}%
  \BibitemOpen
  \bibfield  {author} {\bibinfo {author} {\bibfnamefont {H.}~\bibnamefont
  {Kobayashi}}, \bibinfo {author} {\bibfnamefont {S.}~\bibnamefont {Tamate}},
  \bibinfo {author} {\bibfnamefont {T.}~\bibnamefont {Nakanishi}}, \bibinfo
  {author} {\bibfnamefont {K.}~\bibnamefont {Sugiyama}}, \ and\ \bibinfo
  {author} {\bibfnamefont {M.}~\bibnamefont {Kitano}},\ }\bibfield  {title}
  {\enquote {\bibinfo {title} {Direct observation of geometric phases using a
  three-pinhole interferometer},}\ }\href {\doibase 10.1103/PhysRevA.81.012104}
  {\bibfield  {journal} {\bibinfo  {journal} {Phys. Rev. A}\ }\textbf {\bibinfo
  {volume} {81}},\ \bibinfo {pages} {012104} (\bibinfo {year}
  {2010})}\BibitemShut {NoStop}%
\bibitem [{\citenamefont {Kobayashi}\ \emph {et~al.}(2011)\citenamefont
  {Kobayashi}, \citenamefont {Tamate}, \citenamefont {Nakanishi}, \citenamefont
  {Sugiyama},\ and\ \citenamefont {Kitano}}]{kobayashi11}%
  \BibitemOpen
  \bibfield  {author} {\bibinfo {author} {\bibfnamefont {H.}~\bibnamefont
  {Kobayashi}}, \bibinfo {author} {\bibfnamefont {S.}~\bibnamefont {Tamate}},
  \bibinfo {author} {\bibfnamefont {T.}~\bibnamefont {Nakanishi}}, \bibinfo
  {author} {\bibfnamefont {K.}~\bibnamefont {Sugiyama}}, \ and\ \bibinfo
  {author} {\bibfnamefont {M.}~\bibnamefont {Kitano}},\ }\bibfield  {title}
  {\enquote {\bibinfo {title} {Observation of geometric phases in quantum
  erasers},}\ }\href {\doibase 10.1143/JPSJ.80.034401} {\bibfield  {journal}
  {\bibinfo  {journal} {Journal of the Physical Society of Japan}\ }\textbf
  {\bibinfo {volume} {80}},\ \bibinfo {pages} {034401} (\bibinfo {year}
  {2011})}\BibitemShut {NoStop}%
\bibitem [{\citenamefont {Lundeen}\ and\ \citenamefont
  {Steinberg}(2009)}]{lundeen09}%
  \BibitemOpen
  \bibfield  {author} {\bibinfo {author} {\bibfnamefont {J.~S.}\ \bibnamefont
  {Lundeen}}\ and\ \bibinfo {author} {\bibfnamefont {A.~M.}\ \bibnamefont
  {Steinberg}},\ }\bibfield  {title} {\enquote {\bibinfo {title} {Experimental
  joint weak measurement on a photon pair as a probe of hardy's paradox},}\
  }\href {\doibase 10.1103/PhysRevLett.102.020404} {\bibfield  {journal}
  {\bibinfo  {journal} {Phys. Rev. Lett.}\ }\textbf {\bibinfo {volume} {102}},\
  \bibinfo {pages} {020404} (\bibinfo {year} {2009})}\BibitemShut {NoStop}%
\bibitem [{\citenamefont {Lundeen}\ \emph {et~al.}(2011)\citenamefont
  {Lundeen}, \citenamefont {Sutherland}, \citenamefont {Patel}, \citenamefont
  {Stewart},\ and\ \citenamefont {Bamber}}]{lundeen11}%
  \BibitemOpen
  \bibfield  {author} {\bibinfo {author} {\bibfnamefont {J.~S.}\ \bibnamefont
  {Lundeen}}, \bibinfo {author} {\bibfnamefont {B.}~\bibnamefont {Sutherland}},
  \bibinfo {author} {\bibfnamefont {A.}~\bibnamefont {Patel}}, \bibinfo
  {author} {\bibfnamefont {C.}~\bibnamefont {Stewart}}, \ and\ \bibinfo
  {author} {\bibfnamefont {C.}~\bibnamefont {Bamber}},\ }\bibfield  {title}
  {\enquote {\bibinfo {title} {Direct measurement of the quantum
  wavefunction},}\ }\href@noop {} {\bibfield  {journal} {\bibinfo  {journal}
  {arXiv preprint arXiv:1112.3575}\ } (\bibinfo {year} {2011})}\BibitemShut
  {NoStop}%
\bibitem [{\citenamefont {Bamber}\ and\ \citenamefont
  {Lundeen}(2014)}]{lundeen14}%
  \BibitemOpen
  \bibfield  {author} {\bibinfo {author} {\bibfnamefont {C.}~\bibnamefont
  {Bamber}}\ and\ \bibinfo {author} {\bibfnamefont {J.~S.}\ \bibnamefont
  {Lundeen}},\ }\bibfield  {title} {\enquote {\bibinfo {title} {Observing
  dirac's classical phase space analog to the quantum state},}\ }\href
  {\doibase 10.1103/PhysRevLett.112.070405} {\bibfield  {journal} {\bibinfo
  {journal} {Phys. Rev. Lett.}\ }\textbf {\bibinfo {volume} {112}},\ \bibinfo
  {pages} {070405} (\bibinfo {year} {2014})}\BibitemShut {NoStop}%
\bibitem [{\citenamefont {Malik}\ \emph {et~al.}(2014)\citenamefont {Malik},
  \citenamefont {Lavery}, \citenamefont {Padgett}, \citenamefont {Boyd},
  \citenamefont {Mirhosseini},\ and\ \citenamefont {Leach}}]{malik2014}%
  \BibitemOpen
  \bibfield  {author} {\bibinfo {author} {\bibfnamefont {M.}~\bibnamefont
  {Malik}}, \bibinfo {author} {\bibfnamefont {M.~P.~J.}\ \bibnamefont
  {Lavery}}, \bibinfo {author} {\bibfnamefont {M.~J.}\ \bibnamefont {Padgett}},
  \bibinfo {author} {\bibfnamefont {R.~W.}\ \bibnamefont {Boyd}}, \bibinfo
  {author} {\bibfnamefont {M.}~\bibnamefont {Mirhosseini}}, \ and\ \bibinfo
  {author} {\bibfnamefont {J.}~\bibnamefont {Leach}},\ }\bibfield  {title}
  {\enquote {\bibinfo {title} {Direct measurement of quantum state
  rotations},}\ }\href@noop {} {\bibfield  {journal} {\bibinfo  {journal}
  {Nature Commun.}\ }\textbf {\bibinfo {volume} {5}},\ \bibinfo {pages} {3115}
  (\bibinfo {year} {2014})}\BibitemShut {NoStop}%
\bibitem [{\citenamefont {Salvail}\ \emph {et~al.}(2013)\citenamefont
  {Salvail}, \citenamefont {Agnew}, \citenamefont {Johnson}, \citenamefont
  {Bolduc}, \citenamefont {Leach},\ and\ \citenamefont {Boyd}}]{salvail13}%
  \BibitemOpen
  \bibfield  {author} {\bibinfo {author} {\bibfnamefont {J.~Z.}\ \bibnamefont
  {Salvail}}, \bibinfo {author} {\bibfnamefont {M.}~\bibnamefont {Agnew}},
  \bibinfo {author} {\bibfnamefont {A.~S.}\ \bibnamefont {Johnson}}, \bibinfo
  {author} {\bibfnamefont {E.}~\bibnamefont {Bolduc}}, \bibinfo {author}
  {\bibfnamefont {J.}~\bibnamefont {Leach}}, \ and\ \bibinfo {author}
  {\bibfnamefont {R.~W}\ \bibnamefont {Boyd}},\ }\bibfield  {title} {\enquote
  {\bibinfo {title} {Full characterization of polarization states of light via
  direct measurement},}\ }\href@noop {} {\bibfield  {journal} {\bibinfo
  {journal} {Nature Photonics}\ }\textbf {\bibinfo {volume} {7}},\ \bibinfo
  {pages} {316} (\bibinfo {year} {2013})}\BibitemShut {NoStop}%
\bibitem [{\citenamefont {Sjöqvist}(2006)}]{sjoqvist06}%
  \BibitemOpen
  \bibfield  {author} {\bibinfo {author} {\bibfnamefont {E.}~\bibnamefont
  {Sjöqvist}},\ }\bibfield  {title} {\enquote {\bibinfo {title} {Geometric
  phase in weak measurements},}\ }\href {\doibase
  https://doi.org/10.1016/j.physleta.2006.06.028} {\bibfield  {journal}
  {\bibinfo  {journal} {Physics Letters A}\ }\textbf {\bibinfo {volume}
  {359}},\ \bibinfo {pages} {187 -- 189} (\bibinfo {year} {2006})}\BibitemShut
  {NoStop}%
\bibitem [{\citenamefont {Brunner}\ \emph {et~al.}(2004)\citenamefont
  {Brunner}, \citenamefont {Scarani}, \citenamefont {Wegm\"uller},
  \citenamefont {Legr\'e},\ and\ \citenamefont {Gisin}}]{brunner04}%
  \BibitemOpen
  \bibfield  {author} {\bibinfo {author} {\bibfnamefont {N.}~\bibnamefont
  {Brunner}}, \bibinfo {author} {\bibfnamefont {V.}~\bibnamefont {Scarani}},
  \bibinfo {author} {\bibfnamefont {M.}~\bibnamefont {Wegm\"uller}}, \bibinfo
  {author} {\bibfnamefont {M.}~\bibnamefont {Legr\'e}}, \ and\ \bibinfo
  {author} {\bibfnamefont {N.}~\bibnamefont {Gisin}},\ }\bibfield  {title}
  {\enquote {\bibinfo {title} {Direct measurement of superluminal group
  velocity and signal velocity in an optical fiber},}\ }\href {\doibase
  10.1103/PhysRevLett.93.203902} {\bibfield  {journal} {\bibinfo  {journal}
  {Phys. Rev. Lett.}\ }\textbf {\bibinfo {volume} {93}},\ \bibinfo {pages}
  {203902} (\bibinfo {year} {2004})}\BibitemShut {NoStop}%
\bibitem [{\citenamefont {Mir}\ \emph {et~al.}(2007)\citenamefont {Mir},
  \citenamefont {Lundeen}, \citenamefont {Mitchell}, \citenamefont {Steinberg},
  \citenamefont {Garretson},\ and\ \citenamefont {Wiseman}}]{mir07}%
  \BibitemOpen
  \bibfield  {author} {\bibinfo {author} {\bibfnamefont {R.}~\bibnamefont
  {Mir}}, \bibinfo {author} {\bibfnamefont {J.~S.}\ \bibnamefont {Lundeen}},
  \bibinfo {author} {\bibfnamefont {M.~W.}\ \bibnamefont {Mitchell}}, \bibinfo
  {author} {\bibfnamefont {A.~M.}\ \bibnamefont {Steinberg}}, \bibinfo {author}
  {\bibfnamefont {J.~L.}\ \bibnamefont {Garretson}}, \ and\ \bibinfo {author}
  {\bibfnamefont {H.~M.}\ \bibnamefont {Wiseman}},\ }\bibfield  {title}
  {\enquote {\bibinfo {title} {A double-slit 'which-way' experiment on the
  complementarity–uncertainty debate},}\ }\href
  {http://stacks.iop.org/1367-2630/9/i=8/a=287} {\bibfield  {journal} {\bibinfo
   {journal} {New Journal of Physics}\ }\textbf {\bibinfo {volume} {9}},\
  \bibinfo {pages} {287} (\bibinfo {year} {2007})}\BibitemShut {NoStop}%
\bibitem [{\citenamefont {Kocsis}\ \emph {et~al.}(2011)\citenamefont {Kocsis},
  \citenamefont {Braverman}, \citenamefont {Ravets}, \citenamefont {Stevens},
  \citenamefont {Mirin}, \citenamefont {Shalm},\ and\ \citenamefont
  {Steinberg}}]{kocsis11}%
  \BibitemOpen
  \bibfield  {author} {\bibinfo {author} {\bibfnamefont {S.}~\bibnamefont
  {Kocsis}}, \bibinfo {author} {\bibfnamefont {B.}~\bibnamefont {Braverman}},
  \bibinfo {author} {\bibfnamefont {S.}~\bibnamefont {Ravets}}, \bibinfo
  {author} {\bibfnamefont {M.~J.}\ \bibnamefont {Stevens}}, \bibinfo {author}
  {\bibfnamefont {R.~P.}\ \bibnamefont {Mirin}}, \bibinfo {author}
  {\bibfnamefont {L.~K.}\ \bibnamefont {Shalm}}, \ and\ \bibinfo {author}
  {\bibfnamefont {A.~M.}\ \bibnamefont {Steinberg}},\ }\bibfield  {title}
  {\enquote {\bibinfo {title} {Observing the average trajectories of single
  photons in a two-slit interferometer},}\ }\href {\doibase
  10.1126/science.1202218} {\bibfield  {journal} {\bibinfo  {journal}
  {Science}\ }\textbf {\bibinfo {volume} {332}},\ \bibinfo {pages} {1170--1173}
  (\bibinfo {year} {2011})}\BibitemShut {NoStop}%
\bibitem [{\citenamefont {Aharonov}\ \emph {et~al.}(2013)\citenamefont
  {Aharonov}, \citenamefont {Popescu}, \citenamefont {Rohrlich},\ and\
  \citenamefont {Skrzypczyk}}]{aharonov13}%
  \BibitemOpen
  \bibfield  {author} {\bibinfo {author} {\bibfnamefont {Y.}~\bibnamefont
  {Aharonov}}, \bibinfo {author} {\bibfnamefont {S.}~\bibnamefont {Popescu}},
  \bibinfo {author} {\bibfnamefont {D.}~\bibnamefont {Rohrlich}}, \ and\
  \bibinfo {author} {\bibfnamefont {P.}~\bibnamefont {Skrzypczyk}},\ }\bibfield
   {title} {\enquote {\bibinfo {title} {Quantum cheshire cats},}\ }\href
  {http://stacks.iop.org/1367-2630/15/i=11/a=113015} {\bibfield  {journal}
  {\bibinfo  {journal} {New Journal of Physics}\ }\textbf {\bibinfo {volume}
  {15}},\ \bibinfo {pages} {113015} (\bibinfo {year} {2013})}\BibitemShut
  {NoStop}%
\bibitem [{\citenamefont {Aharonov}\ and\ \citenamefont
  {Casher}(1984)}]{casher84}%
  \BibitemOpen
  \bibfield  {author} {\bibinfo {author} {\bibfnamefont {Y.}~\bibnamefont
  {Aharonov}}\ and\ \bibinfo {author} {\bibfnamefont {A.}~\bibnamefont
  {Casher}},\ }\bibfield  {title} {\enquote {\bibinfo {title} {Topological
  quantum effects for neutral particles},}\ }\href {\doibase
  10.1103/PhysRevLett.53.319} {\bibfield  {journal} {\bibinfo  {journal} {Phys.
  Rev. Lett.}\ }\textbf {\bibinfo {volume} {53}},\ \bibinfo {pages} {319--321}
  (\bibinfo {year} {1984})}\BibitemShut {NoStop}%
\bibitem [{\citenamefont {He}\ and\ \citenamefont {McKellar}(1993)}]{he93}%
  \BibitemOpen
  \bibfield  {author} {\bibinfo {author} {\bibfnamefont {X.-G.}\ \bibnamefont
  {He}}\ and\ \bibinfo {author} {\bibfnamefont {B.~H.~J.}\ \bibnamefont
  {McKellar}},\ }\bibfield  {title} {\enquote {\bibinfo {title} {Topological
  phase due to electric dipole moment and magnetic monopole interaction},}\
  }\href {\doibase 10.1103/PhysRevA.47.3424} {\bibfield  {journal} {\bibinfo
  {journal} {Phys. Rev. A}\ }\textbf {\bibinfo {volume} {47}},\ \bibinfo
  {pages} {3424--3425} (\bibinfo {year} {1993})}\BibitemShut {NoStop}%
\bibitem [{\citenamefont {Wilkens}(1994)}]{wilkens94}%
  \BibitemOpen
  \bibfield  {author} {\bibinfo {author} {\bibfnamefont {M.}~\bibnamefont
  {Wilkens}},\ }\bibfield  {title} {\enquote {\bibinfo {title} {Quantum phase
  of a moving dipole},}\ }\href {\doibase 10.1103/PhysRevLett.72.5} {\bibfield
  {journal} {\bibinfo  {journal} {Phys. Rev. Lett.}\ }\textbf {\bibinfo
  {volume} {72}},\ \bibinfo {pages} {5--8} (\bibinfo {year}
  {1994})}\BibitemShut {NoStop}%
\bibitem [{\citenamefont {Aharonov}\ and\ \citenamefont
  {Vaidman}(2008)}]{aharonov08}%
  \BibitemOpen
  \bibfield  {author} {\bibinfo {author} {\bibfnamefont {Y.}~\bibnamefont
  {Aharonov}}\ and\ \bibinfo {author} {\bibfnamefont {L.}~\bibnamefont
  {Vaidman}},\ }\enquote {\bibinfo {title} {The two-state vector formalism: An
  updated review},}\ in\ \href {\doibase 10.1007/978-3-540-73473-4_13} {\emph
  {\bibinfo {booktitle} {Time in Quantum Mechanics}}},\ \bibinfo {editor}
  {edited by\ \bibinfo {editor} {\bibfnamefont {J.G.}\ \bibnamefont {Muga}},
  \bibinfo {editor} {\bibfnamefont {R.~Sala}\ \bibnamefont {Mayato}}, \ and\
  \bibinfo {editor} {\bibfnamefont {{\'I}.L.}\ \bibnamefont {Egusquiza}}}\
  (\bibinfo  {publisher} {Springer Berlin Heidelberg},\ \bibinfo {address}
  {Berlin, Heidelberg},\ \bibinfo {year} {2008})\ pp.\ \bibinfo {pages}
  {399--447}\BibitemShut {NoStop}%
\bibitem [{\citenamefont {Aharonov}\ \emph {et~al.}(2010)\citenamefont
  {Aharonov}, \citenamefont {Popescu},\ and\ \citenamefont
  {Tollaksen}}]{aharonov10}%
  \BibitemOpen
  \bibfield  {author} {\bibinfo {author} {\bibfnamefont {Y.}~\bibnamefont
  {Aharonov}}, \bibinfo {author} {\bibfnamefont {S.}~\bibnamefont {Popescu}}, \
  and\ \bibinfo {author} {\bibfnamefont {J.}~\bibnamefont {Tollaksen}},\
  }\bibfield  {title} {\enquote {\bibinfo {title} {A time-symmetric formulation
  of quantum mechanics},}\ }\href@noop {} {\bibfield  {journal} {\bibinfo
  {journal} {Physics Today}\ }\textbf {\bibinfo {volume} {63}},\ \bibinfo
  {pages} {27} (\bibinfo {year} {2010})}\BibitemShut {NoStop}%
\bibitem [{\citenamefont {Parrott}(2009)}]{parrott09}%
  \BibitemOpen
  \bibfield  {author} {\bibinfo {author} {\bibfnamefont {S.}~\bibnamefont
  {Parrott}},\ }\bibfield  {title} {\enquote {\bibinfo {title} {What do
  quantum" weak" measurements actually measure?}}\ }\href@noop {} {\bibfield
  {journal} {\bibinfo  {journal} {arXiv preprint arXiv:0908.0035}\ } (\bibinfo
  {year} {2009})}\BibitemShut {NoStop}%
\bibitem [{\citenamefont {Svensson}(2013{\natexlab{a}})}]{svensson13}%
  \BibitemOpen
  \bibfield  {author} {\bibinfo {author} {\bibfnamefont {B.~E.~Y.}\
  \bibnamefont {Svensson}},\ }\bibfield  {title} {\enquote {\bibinfo {title}
  {What is a quantum-mechanical ``weak value'' the value of?}}\ }\href
  {\doibase 10.1007/s10701-013-9740-6} {\bibfield  {journal} {\bibinfo
  {journal} {Foundations of Physics}\ }\textbf {\bibinfo {volume} {43}},\
  \bibinfo {pages} {1193--1205} (\bibinfo {year}
  {2013}{\natexlab{a}})}\BibitemShut {NoStop}%
\bibitem [{\citenamefont {Svensson}(2013{\natexlab{b}})}]{svensson13a}%
  \BibitemOpen
  \bibfield  {author} {\bibinfo {author} {\bibfnamefont {B.}~\bibnamefont
  {Svensson}},\ }\bibfield  {title} {\enquote {\bibinfo {title} {Pedagogical
  review of quantum measurement theory with an emphasis on weak
  measurements},}\ }\href@noop {} {\bibfield  {journal} {\bibinfo  {journal}
  {Quanta}\ }\textbf {\bibinfo {volume} {2}},\ \bibinfo {pages} {18--49}
  (\bibinfo {year} {2013}{\natexlab{b}})}\BibitemShut {NoStop}%
\bibitem [{\citenamefont {Sokolovski}(2015)}]{sokolovski15}%
  \BibitemOpen
  \bibfield  {author} {\bibinfo {author} {\bibfnamefont {D.}~\bibnamefont
  {Sokolovski}},\ }\bibfield  {title} {\enquote {\bibinfo {title} {The meaning
  of “anomalous weak values” in quantum and classical theories},}\ }\href
  {\doibase https://doi.org/10.1016/j.physleta.2015.02.018} {\bibfield
  {journal} {\bibinfo  {journal} {Physics Letters A}\ }\textbf {\bibinfo
  {volume} {379}},\ \bibinfo {pages} {1097 -- 1101} (\bibinfo {year}
  {2015})}\BibitemShut {NoStop}%
\bibitem [{\citenamefont {Matzkin}(2019)}]{matzkin19}%
  \BibitemOpen
  \bibfield  {author} {\bibinfo {author} {\bibfnamefont {A.}~\bibnamefont
  {Matzkin}},\ }\bibfield  {title} {\enquote {\bibinfo {title} {Weak values and
  quantum properties},}\ }\href {\doibase 10.1007/s10701-019-00245-3}
  {\bibfield  {journal} {\bibinfo  {journal} {Foundations of Physics}\ }\textbf
  {\bibinfo {volume} {49}},\ \bibinfo {pages} {298--316} (\bibinfo {year}
  {2019})}\BibitemShut {NoStop}%
\bibitem [{\citenamefont {Dressel}\ \emph {et~al.}(2014)\citenamefont
  {Dressel}, \citenamefont {Malik}, \citenamefont {Miatto}, \citenamefont
  {Jordan},\ and\ \citenamefont {Boyd}}]{dressel14a}%
  \BibitemOpen
  \bibfield  {author} {\bibinfo {author} {\bibfnamefont {J.}~\bibnamefont
  {Dressel}}, \bibinfo {author} {\bibfnamefont {M.}~\bibnamefont {Malik}},
  \bibinfo {author} {\bibfnamefont {F.~M.}\ \bibnamefont {Miatto}}, \bibinfo
  {author} {\bibfnamefont {A.~N.}\ \bibnamefont {Jordan}}, \ and\ \bibinfo
  {author} {\bibfnamefont {R.~W.}\ \bibnamefont {Boyd}},\ }\bibfield  {title}
  {\enquote {\bibinfo {title} {Colloquium: Understanding quantum weak values:
  Basics and applications},}\ }\href {\doibase 10.1103/RevModPhys.86.307}
  {\bibfield  {journal} {\bibinfo  {journal} {Rev. Mod. Phys.}\ }\textbf
  {\bibinfo {volume} {86}},\ \bibinfo {pages} {307--316} (\bibinfo {year}
  {2014})}\BibitemShut {NoStop}%
\bibitem [{\citenamefont {Denkmayr}\ \emph {et~al.}(2014)\citenamefont
  {Denkmayr}, \citenamefont {Geppert}, \citenamefont {Sponar}, \citenamefont
  {Lemmel}, \citenamefont {Matzkin}, \citenamefont {Tollaksen},\ and\
  \citenamefont {Hasegawa}}]{denkmayr14}%
  \BibitemOpen
  \bibfield  {author} {\bibinfo {author} {\bibfnamefont {T.}~\bibnamefont
  {Denkmayr}}, \bibinfo {author} {\bibfnamefont {H.}~\bibnamefont {Geppert}},
  \bibinfo {author} {\bibfnamefont {S.}~\bibnamefont {Sponar}}, \bibinfo
  {author} {\bibfnamefont {H.}~\bibnamefont {Lemmel}}, \bibinfo {author}
  {\bibfnamefont {A.}~\bibnamefont {Matzkin}}, \bibinfo {author} {\bibfnamefont
  {J.}~\bibnamefont {Tollaksen}}, \ and\ \bibinfo {author} {\bibfnamefont
  {Y.}~\bibnamefont {Hasegawa}},\ }\bibfield  {title} {\enquote {\bibinfo
  {title} {Observation of a quantum cheshire cat in a matter-wave
  interferometer experiment},}\ }\href@noop {} {\bibfield  {journal} {\bibinfo
  {journal} {Nature communications}\ }\textbf {\bibinfo {volume} {5}},\
  \bibinfo {pages} {4492} (\bibinfo {year} {2014})}\BibitemShut {NoStop}%
\bibitem [{\citenamefont {Ashby}\ \emph {et~al.}(2016)\citenamefont {Ashby},
  \citenamefont {Schwarz},\ and\ \citenamefont {Schlosshauer}}]{ashby16}%
  \BibitemOpen
  \bibfield  {author} {\bibinfo {author} {\bibfnamefont {J.~M.}\ \bibnamefont
  {Ashby}}, \bibinfo {author} {\bibfnamefont {P.~D.}\ \bibnamefont {Schwarz}},
  \ and\ \bibinfo {author} {\bibfnamefont {M.}~\bibnamefont {Schlosshauer}},\
  }\bibfield  {title} {\enquote {\bibinfo {title} {Observation of the quantum
  paradox of separation of a single photon from one of its properties},}\
  }\href {\doibase 10.1103/PhysRevA.94.012102} {\bibfield  {journal} {\bibinfo
  {journal} {Phys. Rev. A}\ }\textbf {\bibinfo {volume} {94}},\ \bibinfo
  {pages} {012102} (\bibinfo {year} {2016})}\BibitemShut {NoStop}%
\bibitem [{\citenamefont {Duprey}\ \emph {et~al.}(2017)\citenamefont {Duprey},
  \citenamefont {Kanjilal}, \citenamefont {Sinha}, \citenamefont {Home},\ and\
  \citenamefont {Matzkin}}]{duprey17}%
  \BibitemOpen
  \bibfield  {author} {\bibinfo {author} {\bibfnamefont {Q.}~\bibnamefont
  {Duprey}}, \bibinfo {author} {\bibfnamefont {S.}~\bibnamefont {Kanjilal}},
  \bibinfo {author} {\bibfnamefont {U.}~\bibnamefont {Sinha}}, \bibinfo
  {author} {\bibfnamefont {D.}~\bibnamefont {Home}}, \ and\ \bibinfo {author}
  {\bibfnamefont {A.}~\bibnamefont {Matzkin}},\ }\bibfield  {title} {\enquote
  {\bibinfo {title} {The quantum cheshire cat effect: Theoretical basis and
  observational implications},}\ }\href@noop {} {\bibfield  {journal} {\bibinfo
   {journal} {arXiv preprint arXiv:1703.02959}\ } (\bibinfo {year}
  {2017})}\BibitemShut {NoStop}%
\bibitem [{\citenamefont {Campos}\ \emph {et~al.}(1989)\citenamefont {Campos},
  \citenamefont {Saleh},\ and\ \citenamefont {Teich}}]{campos89}%
  \BibitemOpen
  \bibfield  {author} {\bibinfo {author} {\bibfnamefont {Richard~A.}\
  \bibnamefont {Campos}}, \bibinfo {author} {\bibfnamefont {Bahaa E.~A.}\
  \bibnamefont {Saleh}}, \ and\ \bibinfo {author} {\bibfnamefont {Malvin~C.}\
  \bibnamefont {Teich}},\ }\bibfield  {title} {\enquote {\bibinfo {title}
  {Quantum-mechanical lossless beam splitter: Su(2) symmetry and photon
  statistics},}\ }\href {\doibase 10.1103/PhysRevA.40.1371} {\bibfield
  {journal} {\bibinfo  {journal} {Phys. Rev. A}\ }\textbf {\bibinfo {volume}
  {40}},\ \bibinfo {pages} {1371--1384} (\bibinfo {year} {1989})}\BibitemShut
  {NoStop}%
\bibitem [{\citenamefont {Campagne-Ibarcq}\ \emph {et~al.}(2014)\citenamefont
  {Campagne-Ibarcq}, \citenamefont {Bretheau}, \citenamefont {Flurin},
  \citenamefont {Auff\`eves}, \citenamefont {Mallet},\ and\ \citenamefont
  {Huard}}]{campagne14}%
  \BibitemOpen
  \bibfield  {author} {\bibinfo {author} {\bibfnamefont {P.}~\bibnamefont
  {Campagne-Ibarcq}}, \bibinfo {author} {\bibfnamefont {L.}~\bibnamefont
  {Bretheau}}, \bibinfo {author} {\bibfnamefont {E.}~\bibnamefont {Flurin}},
  \bibinfo {author} {\bibfnamefont {A.}~\bibnamefont {Auff\`eves}}, \bibinfo
  {author} {\bibfnamefont {F.}~\bibnamefont {Mallet}}, \ and\ \bibinfo {author}
  {\bibfnamefont {B.}~\bibnamefont {Huard}},\ }\bibfield  {title} {\enquote
  {\bibinfo {title} {Observing interferences between past and future quantum
  states in resonance fluorescence},}\ }\href {\doibase
  10.1103/PhysRevLett.112.180402} {\bibfield  {journal} {\bibinfo  {journal}
  {Phys. Rev. Lett.}\ }\textbf {\bibinfo {volume} {112}},\ \bibinfo {pages}
  {180402} (\bibinfo {year} {2014})}\BibitemShut {NoStop}%
\bibitem [{\citenamefont {Hauser}\ \emph {et~al.}(1974)\citenamefont {Hauser},
  \citenamefont {Neuwirth},\ and\ \citenamefont {Thesen}}]{hauser74}%
  \BibitemOpen
  \bibfield  {author} {\bibinfo {author} {\bibfnamefont {U.}~\bibnamefont
  {Hauser}}, \bibinfo {author} {\bibfnamefont {W.}~\bibnamefont {Neuwirth}}, \
  and\ \bibinfo {author} {\bibfnamefont {N.}~\bibnamefont {Thesen}},\
  }\bibfield  {title} {\enquote {\bibinfo {title} {Time-dependent modulation of
  the probability amplitude of single photons},}\ }\href {\doibase
  https://doi.org/10.1016/0375-9601(74)90667-7} {\bibfield  {journal} {\bibinfo
   {journal} {Physics Letters A}\ }\textbf {\bibinfo {volume} {49}},\ \bibinfo
  {pages} {57 -- 58} (\bibinfo {year} {1974})}\BibitemShut {NoStop}%
\bibitem [{\citenamefont {Weber}\ \emph {et~al.}(2014)\citenamefont {Weber},
  \citenamefont {Chantasri}, \citenamefont {Dressel}, \citenamefont {Jordan},
  \citenamefont {Murch},\ and\ \citenamefont {Siddiqi}}]{weber14}%
  \BibitemOpen
  \bibfield  {author} {\bibinfo {author} {\bibfnamefont {S.~J.}\ \bibnamefont
  {Weber}}, \bibinfo {author} {\bibfnamefont {A.}~\bibnamefont {Chantasri}},
  \bibinfo {author} {\bibfnamefont {J.}~\bibnamefont {Dressel}}, \bibinfo
  {author} {\bibfnamefont {A.~N.}\ \bibnamefont {Jordan}}, \bibinfo {author}
  {\bibfnamefont {K.~W.}\ \bibnamefont {Murch}}, \ and\ \bibinfo {author}
  {\bibfnamefont {I.}~\bibnamefont {Siddiqi}},\ }\bibfield  {title} {\enquote
  {\bibinfo {title} {Mapping the optimal route between two quantum states},}\
  }\href@noop {} {\bibfield  {journal} {\bibinfo  {journal} {Nature}\ }\textbf
  {\bibinfo {volume} {511}},\ \bibinfo {pages} {570--573} (\bibinfo {year}
  {2014})}\BibitemShut {NoStop}%
\end{thebibliography}%

\end{document}